\documentclass[fleqn,usenatbib]{mnras}
\usepackage{enumerate}
\usepackage[T1]{fontenc}
\usepackage{ae,aecompl}
\usepackage{indentfirst}
\usepackage{graphicx}
\usepackage{amsmath}
\usepackage{amssymb}
\usepackage{color}
\usepackage{mathptmx}
\usepackage{epsfig}
\usepackage{wasysym}
\usepackage{xfrac}
\usepackage{pifont}
\usepackage[table,xcdraw]{xcolor}
\usepackage{booktabs}
\usepackage{multirow}
\usepackage{comment}
\usepackage{longtable}
\usepackage[normalem]{ulem}
\usepackage{subcaption}
\usepackage{newtxtext,newtxmath}
\usepackage[T1]{fontenc}
\newcommand{\decals}[0]{DECaLS }
\newcommand{\msun}[0]{M$_\odot$}
\newcommand{\statmorph}[0]{{\textsc{Statmorph}} }
\DeclareRobustCommand{\VAN}[3]{#2}
\let\VANthebibliography\thebibliography
\def\thebibliography{\DeclareRobustCommand{\VAN}[3]{##3}\VANthebibliography}

\usepackage{graphicx}	% Including figure files
\usepackage{amsmath}	% Advanced maths commands

\title[Star formation exists in all early-type galaxies]{Star formation exists in all early-type galaxies -- evidence from ubiquitous structure in UV images}

\author[D. Pandey et al.]{
Divya Pandey$^{1}$\thanks{E-mail: divyap@aries.res.in, divyapandey1212@gmail.com
}, Sugata Kaviraj$^{2}$\thanks{E-mail:s.kaviraj@herts.ac.uk}, Kanak Saha$^{3}$, Saurabh Sharma$^{1}$\\
$^{1}$Aryabhatta Research Institute of Observational Sciences, Manora Peak, Nainital 263 002, India\\
$^{2}$Centre for Astrophysics Research, University of Hertfordshire, College Lane, Hatfield AL10 9AB, UK\\
$^{3}$Inter-University Centre for Astronomy \& Astrophysics, Postbag 4, Ganeshkhind, Pune 411 007, India
}

\begin{document}
\label{firstpage}
\pagerange{\pageref{firstpage}--\pageref{lastpage}}
\maketitle

\begin{abstract}
Recent surveys have demonstrated the widespread presence of UV emission in early-type galaxies (ETGs), suggesting the existence of star formation in many of these systems. However, potential UV contributions from old and young stars, together with model uncertainties, makes it challenging to confirm the presence of young stars using integrated photometry alone. This is particularly true in ETGs that are fainter in the UV and have red UV-optical colours. An unambiguous way of disentangling the source of the UV is to look for structure in UV images. Optical images of ETGs, which are dominated by old stars, are smooth and devoid of structure. If the UV is also produced by these old stars, then the UV images will share this smoothness, while, if driven by young stars, they will exhibit significant structure. We compare the UV and optical morphologies of 32 ETGs (93 per cent of which are at $z<0.03$) using quantitative parameters (concentration, asymmetry, clumpiness and the S\'ersic index), calculated via deep UV and optical images with similar resolution. Regardless of stellar mass, UV-optical colour or the presence of interactions, the asymmetry and clumpiness of ETGs is significantly larger (often by several orders of magnitudes) in the UV than in the optical, while the UV S\'ersic indices are typically lower than their optical counterparts. The ubiquitous presence of structure demonstrates that the UV flux across our entire ETG sample is dominated by young stars and indicates that star formation exists in all ETGs in the nearby Universe.   
\end{abstract}

%It is noteworthy that this is also true of 


%%%%%%%%%%%%%%%%%%%%%%%%%%%%%%%%%%%%%%%%%%%%%%%%%%

\section{Introduction}

Massive early-type galaxies (ETGs), i.e. elliptical and S0 systems, dominate the stellar mass density in the nearby Universe \citep[e.g.][]{Baldry2004}, making them fundamental to a complete understanding of galaxy evolution. Classical models of ETG evolution, largely underpinned by their optical properties -- e.g., red optical colours with low intrinsic scatter \citep[e.g.][]{Gladders1998,Kaviraj2005} and high alpha-to-iron ratios \citep[e.g.][]{Thomas2005} -- have postulated that these galaxies form in rapid starbursts at high redshift ($z>2$), followed by passive ageing thereafter \citep[e.g.][]{Larson1974,Chiosi2002}. In this `monolithic' formation scenario, the stellar populations of ETGs in the nearby Universe are, by construction, purely old. 

While old stellar populations are generally expected to be faint in the UV, some ETGs (typically bright systems in dense environments) show an upturn in flux \citep{Burstein1988,Oconnell1999} between the near-UV (NUV; 2500\AA) and the far-UV (FUV; 1500\AA). In the context of the classical model, the flux in these `UV upturn' systems has been postulated to originate from evolved metal-rich horizontal branch stars \citep[e.g.][]{Yi2008}. An alternative source of UV flux in old stellar populations could come from hot subdwarf stars in binary systems, which are expected to be in place by $z\sim1$ \citep[e.g.][]{Han2007}. However, the predicted redshift evolution of the FUV to NUV slope in such models does not fit observational data \citep{Ree2007}, making them less likely to be the dominant source of old UV emission in ETGs.  

The unexpected recent discovery, via GALEX \citep[e.g.][]{Martin2005} observations, of the widespread presence of UV sources in ETGs \citep[e.g.][]{Yi2005,Kaviraj2007}, has challenged these longstanding models for their formation. While massive ETGs in large surveys like the SDSS \citep[e.g.][]{Abazajian2009} exhibit the well-known optical colour-magnitude relation with low intrinsic scatter, the same galaxies show almost 6 magnitudes of spread in UV-optical colours \citep[e.g.][]{Kaviraj2007,Schawinski2007}. Comparison to the most extreme nearby UV upturn ETGs -- in which the UV flux is assumed to be driven by old stars -- shows that around a third of massive ETGs have bluer UV-optical colours than even the strongest UV upturn galaxies. This suggests that these ETGs are likely to have experienced star-formation events within the last $\sim$0.5 Gyrs, with the young stars accounting for at least a few percent of the stellar mass of the galaxy \citep[e.g.][]{Kaviraj2007}. 

Interestingly, this large spread in the rest-frame UV-optical colours of massive ETGs remains virtually unchanged at intermediate redshift \citep[$z\sim0.5$, e.g.][]{Kaviraj2008}. At this epoch, the Universe is too young for the UV-producing horizontal-branch to be in place (although binary hot subdwarfs may appear at earlier epochs), further suggesting a persistent presence of young stars in the ETG population. {\color{black}It is worth noting that the existence of a sub-population of blue, star-forming ETGs in the nearby Universe, which show strong structural evidence for merger-driven star formation, supports the idea that not all the stellar mass in ETG population was formed in the early Universe \citep[e.g.][]{George2023a,George2023b}}. Finally, these UV results are also supported by studies that have found strong emission lines (e.g. H$\alpha$$\lambda 6563\AA$ and O[III]$\lambda 5007\AA$) in nearby massive ETGs \citep[e.g.][]{Fukugita2004,Dhiwaretal2023} and evidence for star formation from core-collapse supernovae \citep[e.g.][]{Irani2022}. 

A strong coincidence between the presence of UV flux and morphological disturbances indicative of (minor) mergers \citep[e.g.][]{Kaviraj2011} suggests that the trigger for the star formation seen in the ETG population are interactions with lower mass galaxies, which bring in the gas that fuels the star formation \citep{Kaviraj2009}. This picture appears consistent with the presence of molecular gas in ETGs \citep[e.g.][]{Davis2015,Williams2023} and the fact that the gas and stars are often kinematically misaligned, suggesting an external origin for the gas \citep[e.g.][]{Combes2007}. Taken together, these studies demonstrate that, rather than being passively-evolving, many massive ETGs continue to build stellar mass at late epochs ($z<1$) through low-level star formation driven by minor mergers, the process that has also been shown to drive their size evolution \citep[e.g.][]{Ryan2012,Newman2012}. 

However, some questions remain. The fraction of ETGs with star formation is often estimated via comparison to the UV-optical colour of strong UV upturn galaxies, in which the UV flux is assumed to come from old stars. The rationale is that, if the ETG in question is bluer than strong UV upturn systems, then some of its UV flux must be contributed by sources other than old stellar populations, i.e., young stars. For example, \citet{Kaviraj2007} use $(NUV-r)=5.5$, the rest-frame colour of NGC 4552, as the blue limit that can be achieved by old stars alone. Since around a third of the ETGs in their study have $(NUV-r)$ colours are bluer than this value, they conclude that the fraction of ETGs that host recent star formation is likely to be at least 30 percent. This value is, by construction, a lower limit, which leads to the question: what fraction of ETGs with $(NUV-r)>5.5$ are likely to host star formation activity, and could a more precise value for the fraction of star-forming ETGs be calculated? While the persistent large scatter in the $(NUV-r)$ colours of ETGs out to $z\sim0.5$ (where old horizontal branch stars do not exist) suggests that the star-forming fraction is probably much larger than 30 percent, a potential contribution to the UV flux from hot subdwarf stars (which can be in place at $z\sim1$) complicates this argument. 

An unambiguous route to disentangling the primary source of the UV flux seen in ETGs is to look for the presence of structure in the UV images. It is well established that the distribution of optical light in ETGs, which is driven by the old stellar population, is smooth and devoid of structure \citep[e.g.][]{devaucouleurs1959}, exhibits relatively high S\'ersic indices \citep[typically greater than 2, e.g.][]{Graham2013} and shows low values of morphological parameters such as asymmetry and clumpiness \citep[e.g.][]{Conselice2003}. If the UV is driven by the same old stars, then the UV images will share the smoothness and lack of structure seen in the optical images. However, since star formation is inherently patchy \citep[e.g.][]{Crockett2011}, if the UV flux in ETGs is driven by young stars, then one expects to see significant structure in the UV images, which should manifest itself as higher values of asymmetry and clumpiness and lower values of the S\'ersic index than what is measured in the optical. Crucially, this method will indicate the presence of young stars irrespective of the strength of the UV flux in the ETG in question. 

This exercise requires deep, high-resolution images of a large sample of ETGs in both the optical and the UV, ideally where the resolution of the optical and UV images are similar. It is also best performed at low redshift, where the structure induced by star formation is well-resolved. The purpose of this paper to perform such a study, using optical and UV images from the Dark Energy Camera Legacy Survey \citep[\decals,][]{2019Dey} and the Ultra Violet Imaging Telescope \citep[UVIT,][]{2017Tandon} respectively. 

Differences in the UV and optical morphologies of ETGs have already been noted by some past studies. For example, \citet{2002Windhorst} find that galaxies that are classified as ETGs in the optical wavelengths can show a variety of morphological structures in the mid-UV that might lead to a different morphological classification in these wavelengths. \citet{Kuchinski_2001} find that  some early-type spiral galaxies have high asymmetry in the FUV, driven by features that are enhanced in the FUV compared to the optical images. \citet{Mager2018} have recently studied the UV morphologies of galaxies of all morphological types (including ETGs) using GALEX images. Notwithstanding the relatively low (6 arcsecond) resolution of GALEX, they find that the measured concentrations and asymmetries of ETGs in the UV resemble those of spiral and peculiar galaxies in the optical, suggesting the presence of disks containing recent star formation in the ETG population. While \citet{Mager2018} do not find significant differences between the clumpiness of ETG images in the UV and optical, this is most likely due to the poor resolution of GALEX when compared to ground and space based optical telescopes. Here, we use deep optical and UV images of ETGs at low redshift which have similar resolution, enabling us to perform a more precise and consistent comparison between the optical and UV morphologies in the ETG population. 

%\citep[e.g.,][]{Kuchinski_2001, 2002Windhorst, Mager2018}.

The plan for this paper is as follows. In Section \ref{sec:data}, we describe the instruments from which our UV and optical images are derived and the selection of a sample of nearby ETGs, using morphological classifications from the Galaxy Zoo citizen-science project \citep{Lintott2011,Willett2013}. In Section \ref{sec:morphs}, we first describe the calculation of morphological parameters (concentration, asymmetry, clumpiness and the S\'ersic index) of our ETGs, in the optical and the UV. We then compare the UV and optical morphologies, as a function of stellar mass and UV-optical colour, and discuss our findings in the context of whether the primary UV sources in ETGs are old or young. We summarise our results in Section \ref{sec:summary}.

%%%%%%%%%%%%%%%%%%%%%%%%%%%%%%%%%%%%%%%%%%%%%%%%%%

\section{Data}
\label{sec:data}

\subsection{UVIT}
We use archival FUV imaging observations from UVIT, on board {\em AstroSat}, for the morphological analysis of our ETGs in the UV. The FUV observations were conducted in the BaF2 ($\rm \lambda_e\, =\, 1541$ \AA ) or CaF2 ($\rm \lambda_e\, =\, 1481$ \AA) broadband filters, over a two-year period between 2016 and 2017. We employ CCDLAB \citep{2017Postma} to reduce, and perform various corrections on, UVIT Level 1 data to create our science-ready images. \textcolor{black}{The data reduction includes flat field correction, drift correction and corrections to mitigate fixed pattern noise. The frames containing a large number of photon counts due to cosmic rays are removed in the process. The corrected orbit-wise images are merged together through a centroid alignment process to create the final science-ready image.} A comprehensive catalogue for the entire data set is under preparation (Piridi et al., under review in ApJS). The UVIT field-of-view (FOV) spans a diameter of 28$^{\prime}$. Here, we only use reduced UVIT images which have effective exposure times greater than 10 ks, which facilitates the study of faint UV structure in our ETGs. The 3$\sigma$/ 5$\sigma$ point-source detection limit of the UVIT FUV observation with the lowest exposure time ($\approx$11 ks) in our sample is 26.0/25.5 magnitudes. However, the majority of our galaxies have UVIT observations with exposure times of 22.7 ks which have a 3$\sigma$/5$\sigma$ detection limit of 26.4/25.9 magnitudes. In comparison, the GALEX Deep Imaging Survey (DIS) observations, which have similar exposure times ($\sim$ 30 ks), reach a 5$\sigma$ detection limit of 24.8 magnitudes in the FUV filter \citep{Bianchi2017}, around one magnitude shallower than UVIT. We note that, even with an increased exposure time with GALEX, the magnitude limit reached by UVIT observations is unattainable, because the confusion limit of GALEX for UV sources lies close to 25 mag. 
%{\color{teal} A similar trend in the detection limit follows for NUV observations of UVIT and GALEX.} \textbf{(I think this is true for both the FUV and NUV observations from GALEX DIS. Check if this true and change this sentence accordingly.)A: Its is true for NUV as well. We do not have corresponding NUV image for 22.7ks FUV observation. Do I calculate for a different field and write values?} 
The full width at half maximum (FWHM) of the UVIT FUV point spread function (PSF) varies between 1.3$^{\prime\prime}$ and 1.5$^{\prime\prime}$ \citep{Tandon_2020}. This is nearly three (four) times better than GALEX FUV (NUV) imaging and is comparable to the spatial resolution of large-scale optical surveys, such as DECaLS.  

\begin{figure*}
\center
\includegraphics[width=2\columnwidth]{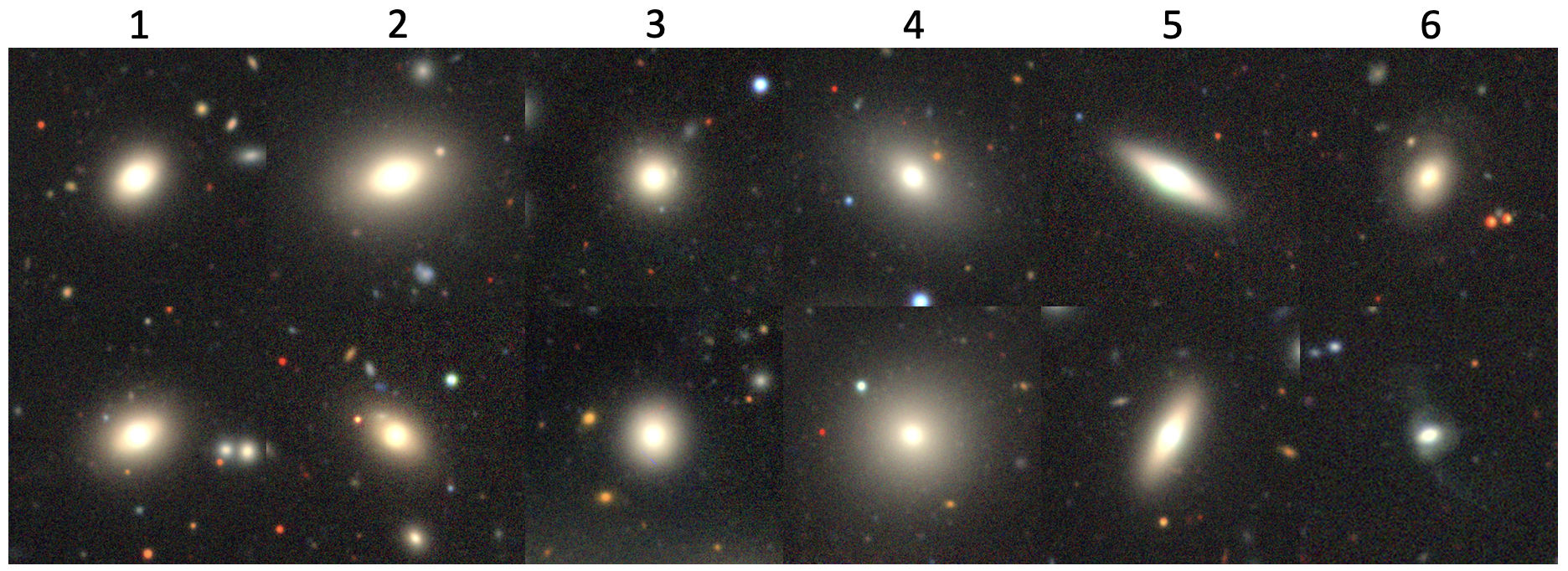}
\caption{\decals images of a random sample of ETGs in our analysis. Columns 1 -- 4 show galaxies classified as ellipticals, column 5 shows galaxies classified as S0s and column 6 shows examples of ETGs that are classified as having signs of an interaction.}
\label{fig:etg_examples}
\end{figure*}

% We estimate the value of background flux and noise in image using random box method. The segmentation maps used to mask sources were created using Source Extractor (ref). 

\subsection{DECaLS and SDSS}

{\color{black} We use sky-subtracted co-added images from \decals, which uses the Dark Energy Camera (DECam) and offers a sky area of $\sim$1400 deg$^2$ in the northern hemisphere in three optical bands ($g$, $r$, and $z$). The median FWHM of the DECaLS PSF is 1.3$^{\prime\prime}$ and the 5$\sigma$ limiting magnitude for a point source within a single frame in $r$-band is $\sim$23.5 magnitudes. Spectroscopic redshifts and stellar masses for our galaxies are taken from the GALEX-SDSS-WISE Legacy Catalog \citep[GSWLC;][]{Salim2016}.  

%SDSS is another large-scale survey conducted in spectroscopic and imaging mode. The imaging observations were carried out in optical and near-infrared broadband filters (u,g,r,i,z). The 5$\sigma$ depth of  SDSS observation in r-band is 22.70 mag. SDSS simultaneously conducts spectroscopic observation of the source within a circular fiber of diameter $\sim$3$^{\prime\prime}$. The redshift values of the elliptical galaxies studied in our sample were retrieved from SDSS.
}

\subsection{A sample of nearby early-type galaxies} 
We use visual morphological classifications, provided by the Galaxy Zoo 2 (GZ2) catalogue \citep{Willett2013}, to assemble a sample of ETGs in the nearby Universe ($z$ $\lesssim$ 0.15). We first select galaxies in GZ2 which have a debiased {\it f\_smoothness} parameter \citep{Willett2013} greater than 0.5. The coordinates of these galaxies are matched with the central coordinates of the FOVs of UVIT pointings, using a radius of 13.5$^{\prime}$, to ascertain whether they have a UVIT observation. 

\begin{figure}
\center
\includegraphics[width=\columnwidth]{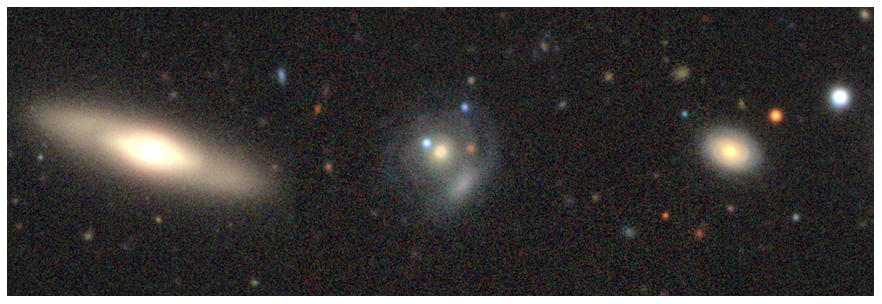}
\caption{\decals images of a random sample of galaxies that are classified as late-types and removed from our analysis.}
\label{fig:ltg_examples2}
\end{figure}

We obtain a 50$^{\prime\prime}$ $\times$ 50$^{\prime\prime}$ FUV cutout for all ETGs that have been observed with UVIT. We run SExtractor \citep{1996Bertin} over the cutouts to select sources that are above a detection limit of 2$\sigma$, with a minimum number of nine pixels required for an object to be identified as a source. This yields an initial sample of 88 galaxies. We then visually inspect the \decals color-composite images of these 88 galaxies to (1) separate our ETGs into elliptical and S0 galaxies (2) identify objects which have signatures of an interaction, such as tidal tails and interacting companions and (3) {\color{black}identify (contaminating) late-type galaxies which show signs of a disc}. We remove these late-type galaxies from our analysis. This leaves a parent sample of ETGs comprising 78 objects, which have stellar masses in the range 10$^{8.85}$ M$_{\odot}$ $<$ M$_{\rm{\star}}$ $<$ 10$^{11.61}$ M$_{\odot}$ and redshifts in the range $0.006<z<0.147$. 

Note that, as described in the next section, to ensure reliable estimation of morphological parameters, we impose additional cuts on the detection signal-to-noise (SNR) and the signal-to-noise per pixel (SNR$_{\rm pixel}$) of our galaxies, which reduces the final sample that is used for the morphological analysis to 32 ETGs. Figure \ref{fig:etg_examples} shows \decals images of a random sample of ETGs in our analysis. Columns 1 -- 4 show galaxies classified as ellipticals, column 5 shows galaxies classified as S0s and column 6 shows examples of ETGs that are classified as having signs of an interaction. {\color{black}Figure \ref{fig:ltg_examples2}} shows example \decals images of galaxies that are classified as late-types and removed from our study. 

Before we begin our analysis, it is worth exploring the biases that may be induced by the detection limits of the UV and optical data used here. Since our goal is to explore the UV and optical morphologies of ETGs in the context of star formation, it is important to estimate the redshifts out to which galaxy populations are complete and the UV red sequence is detectable in these datasets. Using complete samples allows us to probe the presence of star formation in objects that span the full range of UV-colours seen in ETGs. This includes UV-red galaxies where star formation is likely to be weak and the degeneracy between old and young stars, in terms which dominates the UV emission, is difficult to break (e.g. in \citet{Kaviraj2007} the UV-red population is considered to be ETGs where $(NUV-r)$ > 5.5). 

We define the redshift at which galaxy populations of a given stellar mass are complete as the redshift at which a purely-old stellar population of a given stellar mass, that forms in an instantaneous burst at $z=2$, is detectable, at the detection limit of the survey in question. We construct this instantaneous burst using the \citet{Yi2003} stellar models, assuming half solar metallicity. This corresponds to the metallicity at the mean stellar mass of our sample ($\sim$10$^{10.3}$ M$_{\odot}$), calculated using the mass-metallicity relation in the nearby Universe \citep[e.g.][]{Panter2008}. Such a purely-old population represents a faintest `limiting' case, since real galaxies, which are not composed uniquely of old stars, are more luminous than this limiting value. If this limiting case is detectable at the depth of a survey, then it is reasonable to conclude that the entire galaxy population at a given stellar mass will also be detectable. 

\begin{figure}
\center
\includegraphics[width=\columnwidth]{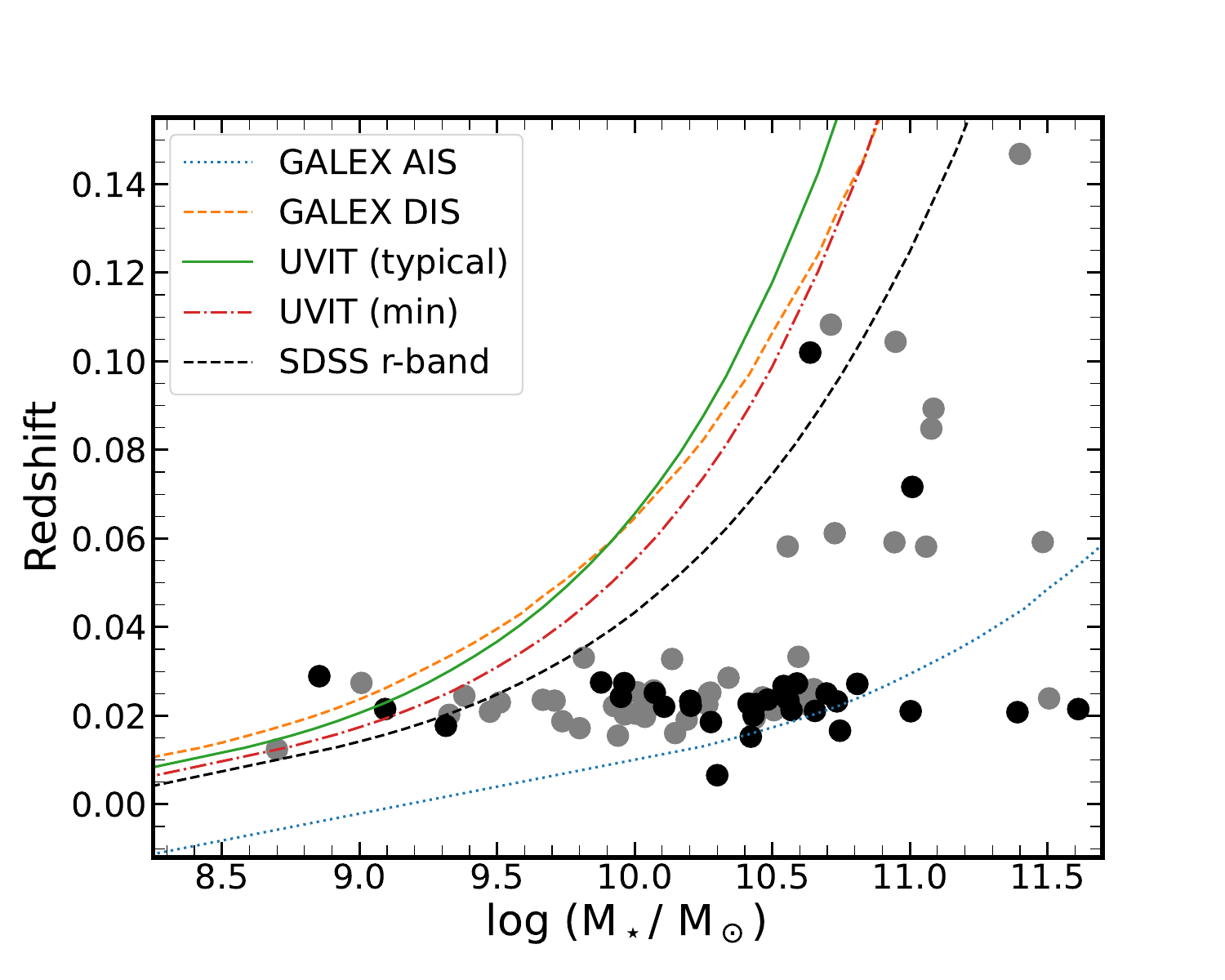}
\caption{Redshifts at which complete galaxy populations can be observed in the SDSS $r$-band, the UVIT observations (red and green) and, for comparison, the GALEX `Deep Imaging Survey' (DIS; orange)  and the GALEX `All-Sky Imaging Survey' (AIS; blue), given the detection limits of the respective observations. The DIS and AIS represent the deepest and shallowest layers of the GALEX surveys respectively. The red dashed-dotted line corresponds to the UVIT observation with the lowest exposure time (10.9 ks), while the green solid line corresponds to a UVIT observation with an exposure time of 22.7 ks (which describes most of our ETGs). The black circles indicate galaxies which are used for our morphological analysis. The grey circles indicate galaxies which are not eventually included in our analysis because they fall outside the signal-to-noise cuts that we impose to ensure robust parameter estimation (see text in Section \ref{sec:morphparams}).}
\label{fig:completeness}
\end{figure}

Figure \ref{fig:completeness} shows the redshifts at which complete galaxy populations can be observed in the SDSS $r$-band, the UVIT observations (red and green) and, for comparison, the GALEX `Deep Imaging Survey' (DIS; orange)  and the GALEX `All-Sky Imaging Survey' (AIS; blue), given the detection limits of the respective observations. The DIS and AIS represent the deepest and shallowest layers of the GALEX surveys respectively. The red dashed-dotted line corresponds to the UVIT observation with the lowest exposure time (10.9 ks), which has a 3$\sigma$ detection limit of 26 magnitudes, while the green solid line corresponds to a UVIT observation with an exposure time of 22.7 ks which has a 3$\sigma$ detection limit of 26.4 magnitudes. The latter describes most of the ETGs in this study. The black circles indicate galaxies which are used for our morphological analysis. The grey circles indicate galaxies which are not eventually included in our analysis because they fall outside the signal-to-noise cuts that we impose to ensure robust parameter estimation (see Section \ref{sec:morphparams} below). This figure demonstrates that the overwhelming majority of our ETGs are in parts of the redshift vs stellar mass space where even purely old stellar populations are visible and galaxy populations will therefore be complete and unbiased. 

\begin{figure*}
    \includegraphics[width=\linewidth]{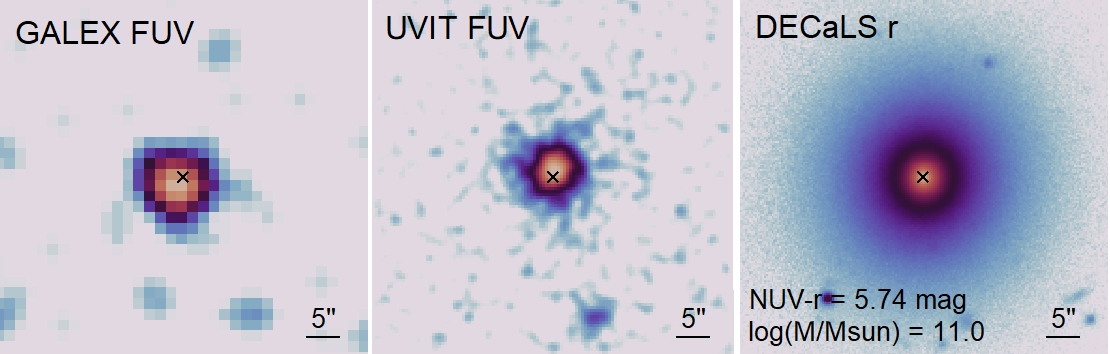}
    \includegraphics[width=\linewidth]{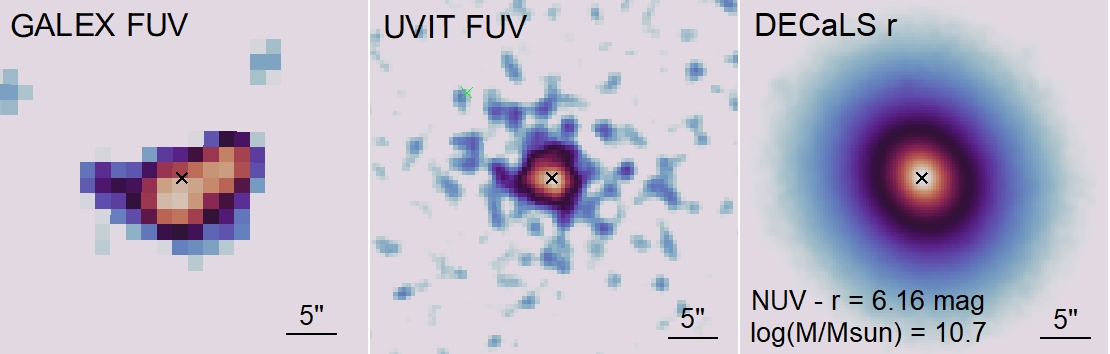}
    \caption{GALEX FUV, UVIT FUV and \decals $r$-band images of two galaxies from our sample. The black crosses represent the optical centroids of each galaxy. The GALEX-SDSS $(NUV-r)$ colour and the stellar mass of each galaxy is indicated on the images. Each panel represents the same region for each galaxy.}
    \label{fig:galaxy_images}
\end{figure*}

%{\color{teal} We use visual morphological classifications provided by the Galaxy Zoo 2 (GZ2) catalogue \citep{Willett2013} to assemble a sample of ETGs in the nearby Universe ($z$ $\lesssim$ 0.1). The morphological classifications provided by GZ2 utilizes SDSS Data Release 7 imaging observations. The project is based on visual structural classification performed by volunteer citizens. We apply a limit on the debiased {\it f\_smoothness} parameter $\geq$ 0.5, to identify elliptical galaxies. Agreeable, with a rather lenient cut, we may include a few late-type galaxies in the sample. Therefore, we overview the morphology of selected galaxies at a later stage. The coordinates of identified early-type galaxies are matched with the central coordinates of the FOV of UVIT observations with a radius of 13.5$^{\prime}$. We obtain a small cutout of 50$^{\prime\prime}$ $\times$ 50$^{\prime\prime}$ from FUV images for all elliptical galaxies observed with UVIT. Further, we run Source Extractor \citep{1996Bertin} over the cutout images to select sources above a detection limit of 2$\sigma$ (check) with minimum number of pixel to be detected as a source as 9 pixels. This leaves us with a sample of 78 galaxies. We visually examine the \decals color composite images of these galaxies to identify interaction signatures such as a tidal tail, galaxies with close neighbors, or late-type galaxies. The detected potential spirals are removed from the sample.} 

%%%%%%%%%%%%%%%%%%%%%%%%%%%%%%%%%%%%%%%%%%%%%%%%%%

\section{UV vs optical morphology of nearby ETGs}
\label{sec:morphs}

\subsection{Measurement of morphological parameters}

\label{sec:morphparams}

{\color{black} We use two well-known approaches, the `CAS' system \citep{2000Conselice,Conselice2003} and the S\'ersic index \citep{Sersic1963} to quantitatively compare the morphology of ETGs in the UV and optical wavebands. We begin by briefly describing these approaches. 

CAS comprises three parameters -  concentration ($C$), asymmetry ($A$) and clumpiness ($S$) - based on a pixel-wise analysis of observed physical features within galaxies. {\color{black}A rich literature has used the CAS methodology, both to separate galaxies of different morphological types \citep[e.g.][]{Bershady_2000,Mager2018,Sazonova2020,Nersesian2023} and to identify interesting sub-populations like merging and interacting galaxies \citep[e.g.][]{Conselice_2003,Lotz2004,Lotz2008,Holwerda2011,Conselice2014,Sazonova2021}.} %The CAS methodology is used repeatedly to identify underlying merger interaction in galaxies or to separate galaxies of different morphological types \citep{Bershady_2000}. 
We briefly describe the CAS parameters below.

The concentration index is calculated by computing the ratio of the radii which enclose 80 and 20 per cent of the total light of the galaxy respectively. This parameter is defined as follows:

\begin{equation}
    \rm \mathit{C} =5 \times log_{10} \left( \frac{\mathit{R}_{80}}{\mathit{R}_{20}} \right)
    \label{eq:c82}
\end{equation}

\noindent where $R_{\rm 80}$ and $R_{\rm 20}$ correspond to the radii enclosing 80 and 20 per cent of the total light of the galaxy respectively. In massive galaxies, the light concentration varies significantly as a function of morphological type \citep{1994Abraham, Conselice2003}. For example, the more strongly peaked flux distribution in massive ETGs leads to larger values of concentration in the optical wavelengths compared to their late-type counterparts \citep{Bershady_2000, Conselice2003}\footnote{In the dwarf regime, however, the concentration values in ETGs tend to be lower and similar to that in their LTG counterparts \citep[e.g.][]{Lazar2024}.}. 

% Numerically, the concentration parameter can be computed using the following relation:
% \begin{equation}
%  C = 5\times\log\frac{r_{80_\%}}{r_{20_\%}}    
% \end{equation}
% here $r_{80_\%}$ and $r_{20_\%}$ are radius at which 80\% and 20\% light is enclosed within the petrosian radius of a galaxy. 

The asymmetry index is obtained by subtracting the original image of the galaxy from a version where the image is rotated by 180$^{\circ}$. This pixel wise subtraction is performed out to 1.5 times the petrosian radius ($R\rm_{petro}$). In massive galaxies, this index, when calculated using optical images, is typically lower in ETGs, which have fewer star-forming regions, than in their late-type counterparts \citep{1996Abraham, Bershady_2000, Conselice2003}. Asymmetry is defined as follows:

\begin{equation}
    \rm \mathit{A}=\frac{\sum_{\mathit{i},\mathit{j}} \mid \mathit{I}_{ij} - \mathit{I}_{ij}^{180} \mid }{\sum_{\mathit{i},\mathit{j}} \mid \mathit{I}_{ij} \mid  } - \mathit{A}_{bgr}
    \label{eq:asy}
\end{equation}

\noindent where $I\rm_{i,j}$ and $I\rm_{i,j}^{180}$ are the pixel values of the original and the rotated images, respectively and $A\rm_{bgr}$ is the asymmetry of the background. 

Finally, the clumpiness parameter (sometimes also called the smoothness parameter) characterises the patchiness of the light distribution within a galaxy. The parameter is calculated by subtracting a smoothed image from the original image of a galaxy, as defined below.

\begin{equation}
    \rm \mathit{S}=\frac{\sum_{\mathit{i},\mathit{j}}  \mathit{I}_{ij} - \mathit{I}_{ij}^{S} }{\sum_{\mathit{i},\mathit{j}} \mathit{I}_{ij}} - \mathit{S}_{bgr}
    \label{eq:s}
\end{equation}

\noindent where $I\rm_{i,j}$ and $I\rm_{i,j}^{S}$ are the pixel values of the original image and its smoothed version, respectively. The smoothed image is obtained using a boxcar filter of width $\sigma$. The $\sigma$ value used is 0.25 $R\rm_{petro}$, as in \citet{Lotz2004}. In massive galaxies, ETGs typically show a fairly smooth light distribution in the optical wavelengths, resulting in lower clumpiness indices than their late-type counterparts \citep{Conselice2003}.  

We employ \statmorph \citep{2019Rodriguez} to compute CAS parameters for our sample of ETGs, using background-subtracted optical $r$-band and UV images from \decals and UVIT respectively. %The images provided to \statmorph were background subtracted. The program is designed to compute the CAS indices within a few (generally 1.5) times the petrosian radius of the galaxy. 
As we are interested in comparing the CAS parameters of ETGs in the UV and optical wavelengths, we measure the parameters within the same aperture (1.5 times the FUV petrosian radius) in both wavebands. The segmentation required by \statmorph to mask the neighbouring sources around the main galaxy are generated using SExtractor for both the UV and \decals observations. %The software also uses a weight-map comprising the value standard deviation at each pixel. 
Weight-maps in the optical are extracted from the archive for the \decals observations, while the FUV weight maps are computed using \textsc{SExtractor}. \statmorph provides SNR$_{\rm pixel}$ and a flag that quantifies the reliability of the CAS parameters. The values for the flag are integers between 0 and 4, which represent `good' to `catastrophic' quality estimates respectively. We find that 46 (32) out of our 78 galaxies have FUV SNR$_{\rm pixel}$ greater than 2.0 (2.5) with flag$_{\rm FUV}$ $=$ 0 or 1. 

% \textbf{The description of the flag is unclear. What values does it take and what do they mean?}}

%\begin{figure}
%    \includegraphics[width=0.99\linewidth]{gal10.jpg}
%    \includegraphics[width=0.99\linewidth]{gal58.jpg}
%    \includegraphics[width=0.99\linewidth]{gal60.jpg}
%    \caption{GALEX FUV, UVIT FUV and \decals $r$-band images of three galaxies from our sample. The black crosses represent the optical centroids of each galaxy. The GALEX-SDSS $(NUV-r)$ color and stellar mass for each galaxy is indicated on the images. %Black dashed regions highlight clumps in optical and FUV images. Each panel represents the same region for each galaxy.}
%    \label{fig:galaxy_images}
%\end{figure}

The SNR and spatial resolution of the images in question play a key role in determining the reliability of CAS measurements \citep[e.g.][]{2000Conselice, Lotz2004, Lotz_2006}. We make several further cuts to ensure the robustness of our morphological parameters. \citet{Lotz2004} show that the deviation in morphological parameters is less than 10 per cent when they are derived from images that have SNR$_{\rm pixel}$ $\geq$ 2. Furthermore, the $A$ and $S$ parameters are relatively unbiased if calculated using images that have physical resolutions better than around 1 kpc. {\color{black} Following \citet{Lotz2004} -- see also \citet{Rodriguez2015} and \citet{Tohill_2021} -- we restrict our study to galaxies which have SNR$_{\rm pixel}$ $\geq$ 2. {\color{black}This cut reduces our original sample to 42 galaxies with SNR$_{\rm pixel}$ greater than 2.} %This cut reduces our original sample to 30 (42) galaxies with SNR$_{\rm pixel}$ greater than 2.5 (2), respectively. 

$R_{\rm 20}$ -- the radius that encloses 20 per cent of the light -- is another criterion that may impact the reliability of the measured CAS parameters. \citet{Lotz_2006} show that the measured parameters may have artificial values if 20 per cent of the light in a galaxy is concentrated within 1 $-$ 2 pixels. \statmorph recommends that $R_{\rm 20}$ is greater than half of the PSF FWHM of the image to ensure that the measured parameters are reliable. We, therefore, further restrict our study to galaxies where $R_{20}$ $>$ 2 pixels ($\sim$ 0.83$^{\prime\prime}$ for UVIT, which is greater than half of the worst PSF FWHM in our UVIT observations, $\sim$ 1.6$^{\prime\prime}$). {\color{black}With this cut, our final sample contains 32 galaxies with SNR$_{\rm pixel}$ greater than 2 (out of which 23 have SNR$_{\rm pixel}$ greater than 2.5).} %With this cut, our final sample contains 23 (32) galaxies with SNR$_{\rm pixel}$ greater than 2.5 (2), respectively. 
This sample spans a very similar mass (10$^{\rm 8.85}$ M$_\odot$ < $M_{\star}$ < 10$^{\rm 11.6}$ M$_\odot$) and redshift {\color{black}($z < 0.102$)} range as the parent ETG sample described in Section \ref{sec:data}. Around 93 per cent of our ETGs reside at redshifts less than 0.03 and the linear resolution achieved by UVIT up to this redshift is better than 1 kpc.  

In Figures~\ref{fig:galaxy_images} and \ref{fig:snrcmp}, we present a comparison between UVIT and GALEX. Figure~\ref{fig:galaxy_images} shows GALEX FUV, UVIT FUV and \decals $r$-band images for two galaxies in our sample. This figure illustrates the superior resolution of UVIT, which enables it to resolve finer structure in UV images compared to GALEX. Figure~\ref{fig:snrcmp} shows a comparison between the integrated UVIT and GALEX FUV SNRs for our final sample, computed in a similar manner to \citet{Mager2018}. The integrated UVIT FUV SNRs of our ETGs are greater than 20, with around 40 per cent of the galaxies detected with UVIT being undetected (or detected with an integrated SNR $<$ 2) in GALEX.

\begin{figure}
    \centering
    \includegraphics[width=0.99\linewidth]{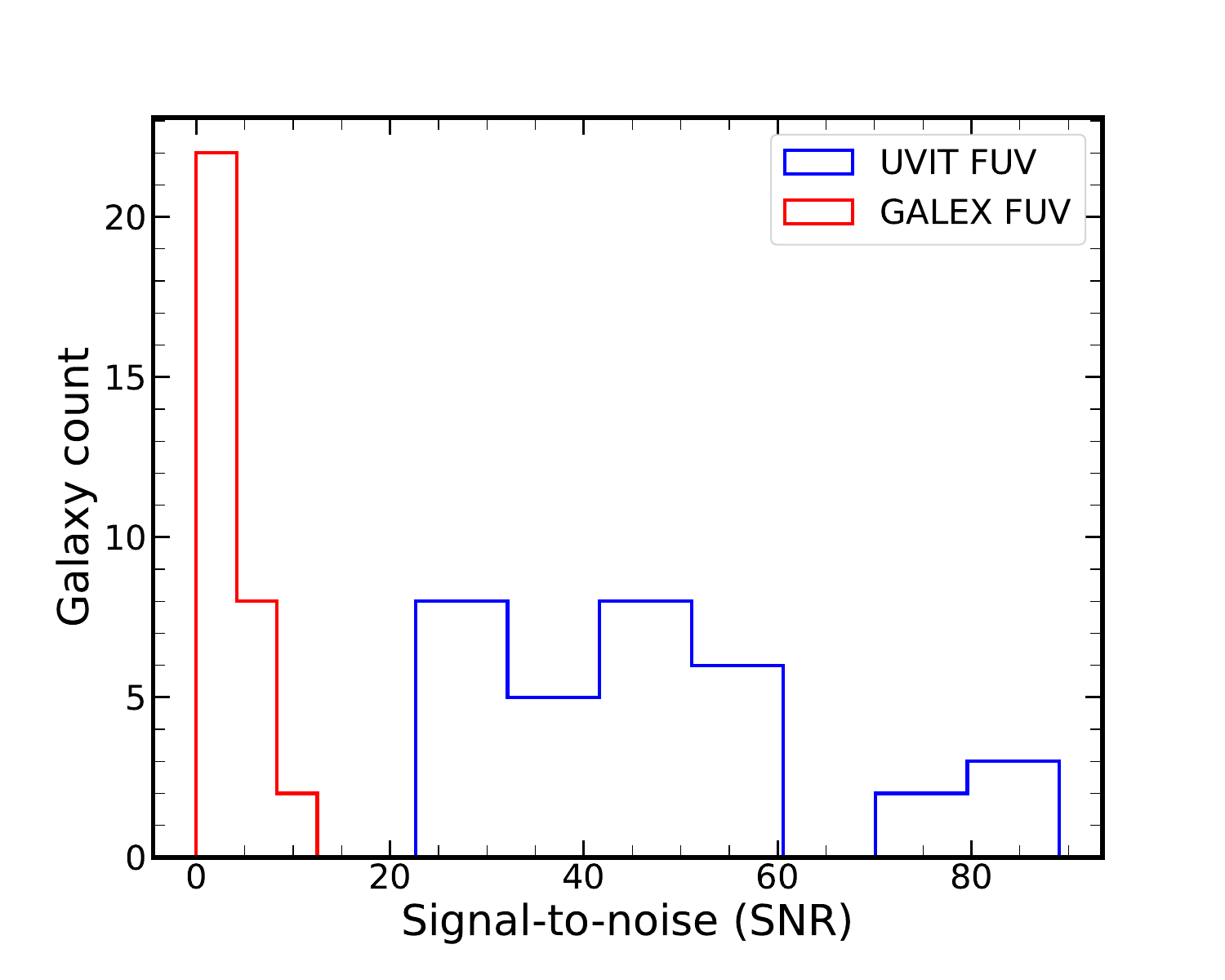}
    \caption{GALEX (red) and UVIT (blue) SNRs, calculated using integrated FUV fluxes, for ETGs in our sample that have SNR$_{\rm pixel}$ $\geq$ 2 and $R_{\rm 20}\, > 2$ pixels (0.83$^{\prime\prime}$).}
    \label{fig:snrcmp}
\end{figure}

{\color{black}The S\'ersic function is commonly-used for parametrising the shape of the light profile of a galaxy \citep{Sersic1963}. The function is described as follows:

\begin{equation}
    I(R) = I_e \exp{ \Bigg\{-b_{\rm n} \Bigg[ \Bigg(\frac{R}{R_{\rm e}}\Bigg)^\frac{1}{n}  - 1 \Bigg]  \Bigg \}},
\end{equation}

where $R_{\rm e}$ is the radius enclosing half the light of the galaxy, $I_{\rm e}$ is the intensity at $R_{\rm e}$ and $b_{\rm n}$ is a function of the S\'ersic index ($n$). We fit one-dimensional S\'ersic profiles to the isophotal distributions in the FUV and $r$-band images of our ETGs, using \textsc{PROFILER} \citep{2016Ciambur}. {\color{black}While the optical $r$-band light profiles in all our ETGs are well-fitted by S\'ersic profiles, the FUV light profiles are more irregular.} %While fitting isophotes to the optical light profiles is straightforward, the FUV light profiles are more irregular. 
This is directly related to the significant amounts of structure that exists in the UV images, as discussed in Section \ref{sec:starformation} below. We, therefore, adopt several steps in the S\'ersic fitting to the FUV images of our ETGs. The FUV images are first convolved with a Gaussian function, which reduces noise in the images and helps determine the position of the centroid of each galaxy. The isophotal fluxes are then measured on the original FUV images. A S\'ersic profile can be successfully derived from the FUV surface-brightness distribution for 19 galaxies. Of these, the errors in the measurement of $n$ ($\Delta\, n$) are less than 0.25 for ten galaxies, and the remaining nine exhibit errors between 0.25 and 1. The FUV light profiles of the remaining ETGs are not smooth enough to fit a S\'ersic function. 

The light profiles of some of our ETGs are shown in Figure~\ref{fig:sersic_fitting}. While the error bars on individual points are plotted, they are too small to be visible. The left-hand column shows examples of ETGs in which the FUV profiles show a significant lack of smoothness (although the mean error in the surface brightness is $\sim$ 0.1 mag/arcsec$^{-2}$). The lack of smoothness of the FUV light profiles is driven by the presence of underlying structure in the FUV images. The right-hand column shows examples of galaxies where S\'ersic functions can be successfully fitted to both the FUV and optical images. Table~\ref{tab:Tab1} lists the coordinates, physical properties and morphological parameters of the ETGs which are included in our analysis. Recall that these are galaxies with FUV SNR$_{\rm pixel}$ $\geq$ 2 and R$_{20}$ $>$ 2 pixels (0.83$^{\prime\prime}$). 

\begin{table*}
    \centering

\caption{Coordinates, morphological and structural parameters and physical properties in the UVIT FUV and \decals $r$-band of our ETGs with FUV SNR$_{\rm pixel}$ $\geq$ 2 and $R_{\rm 20}$ $>$ 2 pixels (0.83$^{\prime\prime}$).}
\begin{tabular}{cccccccccccccc}
%\begin{tabular}{rrrrrrrrrrrr}
\hline\hline
  \multicolumn{1}{c}{ID} &
  \multicolumn{1}{c}{RA (J2000)} &
  \multicolumn{1}{c}{Dec (J2000)} &
  \multicolumn{1}{c}{$z$} &
  \multicolumn{1}{c}{$C_{\rm FUV}$} &
  \multicolumn{1}{c}{$A_{\rm FUV}$} &
  \multicolumn{1}{c}{$S_{\rm FUV}$} &
  \multicolumn{1}{c}{$C_{\rm r}$} &
  \multicolumn{1}{c}{$A_{\rm r}$} &
  \multicolumn{1}{c}{$S_{\rm r}$} &
  \multicolumn{1}{c}{n$_{\rm FUV}$} &
  \multicolumn{1}{c}{n$_{\rm r}$} &
  \multicolumn{1}{c}{$NUV-r$} &
  \multicolumn{1}{c}{log ($M_\star$/ \msun)} \\
\hline
1 & 12:29:59.10 & +12:20:55.2 & 0.007 & 2.24 & 0.507 & 0.163 & 2.778 & 0.099 & 0.002 & 1.37 $\pm$ 0.09 & 2.87 $\pm$ 0.02 & 4.89 & 10.3\\
  2 & 12:59:19.87 & +28:05:03.4 & 0.015 & 2.59 & 0.361 & 0.221 & 2.805 & 0.034 & 0.013 &  & 3.10 $\pm$ 0.09 & 5.71 & 10.42\\
  3 & 13:00:54.45 & +28:00:27.4 & 0.017 & 3.15 & 0.19 & 0.186 & 3.05 & 0.014 & -0.002 &    & 3.18 $\pm$ 0.01 & 5.91 & 10.75\\
  4 & 13:00:09.14 & +27:51:59.3 & 0.018 & 2.77 & 0.235 & 0.122 & 2.576 & 0.158 & 0.023 & 2.02 $\pm$ 0.16 & 1.46 $\pm$ 0.03 & 2.89 & 9.31\\
  5 & 13:02:21.52 & +28:13:50.8 & 0.019 & 2.98 & 0.2 & 0.483 & 3.224 & 0.031 & 0.043 &    & 2.68 $\pm$ 0.01 & 5.8 & 10.28\\
  6 & 11:42:59.06 & +20:05:12.7 & 0.02 & 2.3 & 0.164 & 0.277 & 3.409 & 0.019 & 0.0 &    & 4.19 $\pm$ 0.13 & 5.71 & 10.43\\
  7 & 11:44:02.16 & +19:56:59.4 & 0.021 & 3.03 & 0.394 & 0.262 & 3.218 & 0.004 & -0.005 & 2.11 $\pm$ 0.09 & 4.9 $\pm$ 0.06 & 5.63 & 11.39\\
  8 & 11:43:56.42 & +19:53:40.5 & 0.021 & 2.69 & 0.406 & 0.208 & 3.119 & 0.023 & -0.003 & 3.28 $\pm$ 0.26 & 3.77 $\pm$ 0.03 & 5.74 & 11.0\\
  9 & 11:44:02.16 & +19:58:18.9 & 0.021 & 2.81 & 0.275 & 0.33 & 3.438 & 0.016 & 0.005 &    & 5.28 $\pm$ 0.06 & 5.61 & 10.65\\
  10 & 13:00:42.76 & +27:58:16.5 & 0.021 & 3.16 & 0.185 & 0.177 & 3.083 & 0.043 & 0.002 &   & 4.09 $\pm$ 0.1 & 5.77 & 10.57\\
  11 & 13:00:08.13 & +27:58:37.0 & 0.022 & 2.95 & 0.167 & 0.139 & 2.989 & 0.014 & -0.002 & 3.59 $\pm$ 0.18 & 4.18 $\pm$ 0.05 & 5.19 & 11.61\\
  12 & 11:43:01.19 & +19:54:35.3 & 0.022 & 2.58 & 0.417 & 0.285 & 2.514 & 0.022 & 0.019 & 2.75 $\pm$ 0.44 & 2.3 $\pm$ 0.07 & 3.53 & 9.09\\
 13 & 11:44:05.75 & +20:14:53.5 & 0.022 & 2.51 & 0.409 & 0.318 & 3.208 & 0.023 & -0.002 & 1.99 $\pm$ 0.79 & 2.46 $\pm$ 0.11 & 4.44 & 10.11\\
 14 & 12:59:43.73 & +27:59:40.9 & 0.022 & 3.51 & 0.178 & 0.19 & 3.027 & 0.03 & 0.018 & 3.22 $\pm$ 0.56 & 3.66 $\pm$ 0.08 & 5.9 & 10.2\\
 15 & 12:59:29.41 & +27:51:00.5 & 0.023 & 2.95 & 0.336 & 0.26 & 3.099 & 0.039 & 0.001 &    & 1.97 $\pm$ 0.04 & 5.82 & 10.44\\
 16 & 13:02:21.66 & +28:15:21.5 & 0.023 & 2.54 & 0.316 & 0.259 & 2.542 & 0.023 & -0.002 &    & 1.11 $\pm$ 0.03 & 5.72 & 10.41\\
 17 & 13:00:48.65 & +28:05:26.6 & 0.023 & 3.39 & 0.234 & 0.238 & 3.256 & 0.012 & -0.001 &    & 2.59 $\pm$ 0.03 & 5.67 & 10.73\\
 18 & 13:00:27.97 & +27:57:21.5 & 0.023 & 2.8 & 0.094 & 0.201 & 3.207 & 0.028 & -0.005 &   & 2.36 $\pm$ 0.14 & 6.14 & 10.2\\
 19 & 12:59:44.41 & +27:54:44.8 & 0.023 & 3.57 & 0.162 & 0.161 & 3.209 & 0.024 & 0.015 &    & 2.02 $\pm$ 0.03 & 5.86 & 10.56\\
 20 & 13:00:40.85 & +27:59:47.8 & 0.024 & 2.39 & 0.219 & 0.213 & 2.914 & 0.016 & -0.005 &    & 1.79 $\pm$ 0.03 & 5.9 & 10.48\\
 21 & 12:59:34.12 & +27:56:48.6 & 0.024 & 3.28 & 0.01 & 0.116 & 3.698 & 0.045 & 0.003 &    & 5.15 $\pm$ 0.15 & 4.93 & 10.55\\
 22 & 13:00:06.39 & +28:00:14.9 & 0.024 & 2.79 & 0.21 & 0.247 & 3.158 & 0.028 & -0.003 & 1.39 $\pm$ 0.16 & 3.0 $\pm$ 0.07 & 5.97 & 9.95\\
 23 & 13:00:39.76 & +27:55:26.2 & 0.025 & 3.34 & 0.249 & 0.472 & 3.59 & 0.011 & -0.004 & 1.73 $\pm$ 0.61 & 2.71 $\pm$ 0.02 & 6.16 & 10.7\\
 24 & 13:00:06.25 & +27:41:07.1 & 0.025 & 2.66 & 0.124 & 0.243 & 3.007 & 0.04 & 0.016 &    & 3.7 $\pm$ 0.05 & 5.44 & 10.07\\
 25 & 12:59:37.91 & +27:54:26.3 & 0.027 & 3.67 & 0.134 & 0.252 & 3.359 & 0.024 & 0.015 & 2.44 $\pm$ 0.87 & 3.38 $\pm$ 0.08 & 5.98 & 10.54\\
 26 & 12:59:56.01 & +28:02:05.0 & 0.027 & 3.14 & 0.285 & 0.257 & 3.009 & 0.019 & 0.022 &    & 2.21 $\pm$ 0.02 & 5.77 & 10.81\\
 27 & 13:00:22.13 & +28:02:49.2 & 0.027 & 2.97 & 0.277 & 0.278 & 3.029 & 0.023 & 0.008 & 3.33 $\pm$ 0.53 & 2.42 $\pm$ 0.03 & 6.02 & 10.59\\
 28 & 13:00:33.36 & +27:49:27.3 & 0.027 & 2.4 & 0.221 & 0.353 & 2.712 & 0.009 & 0.007 &    & 1.72 $\pm$ 0.03 & 4.84 & 9.96\\
 29 & 12:59:54.86 & +27:47:45.6 & 0.028 & 2.22 & 0.233 & 0.242 & 2.95 & 0.018 & -0.008 &    & 2.32 $\pm$ 0.09 & 4.58 & 9.88\\
 30 & 13:00:08.06 & +27:46:23.9 & 0.029 & 2.68 & 0.249 & 0.149 & 2.748 & 0.175 & 0.011 & 2.1 $\pm$ 0.13 & 1.99 $\pm$ 0.07 & 2.55 & 8.85\\
 31 & 14:53:45.02 & +18:34:06.1 & 0.072 & 1.95 & 0.289 & 0.276 & 3.029 & 0.035 & -0.004 &   & 2.73 $\pm$ 0.07 & 5.55 & 11.01\\
 32 & 12:19:14.58 & +29:51:21.8 & 0.102 & 2.4 & 0.201 & 0.141 & 2.716 & 0.03 & 0.053 & 1.41 $\pm$ 0.15 & 2.69 $\pm$ 0.08 & 3.0 & 10.64\\

\hline\end{tabular}
\label{tab:Tab1}
\end{table*}

%{\bf It would be worth adding error bars to individual points in your figure to make this clear and mention this in the text above. However, I dont know if that will make the plots too noisy so if you think its not needed then thats fine.} {\color{teal} I have added error bars, since we are showing 10 orders of magnitude range in each panel, the error bars are not visible properly - \textbf{I have added a new Figure~\ref{fig:new_sersic} which shows FUV and r-band in different panels. The error bars are more prominent here. Should we keep the new figure in the paper?}}

\begin{figure*}
    \centering
    \includegraphics[width=0.9\linewidth]{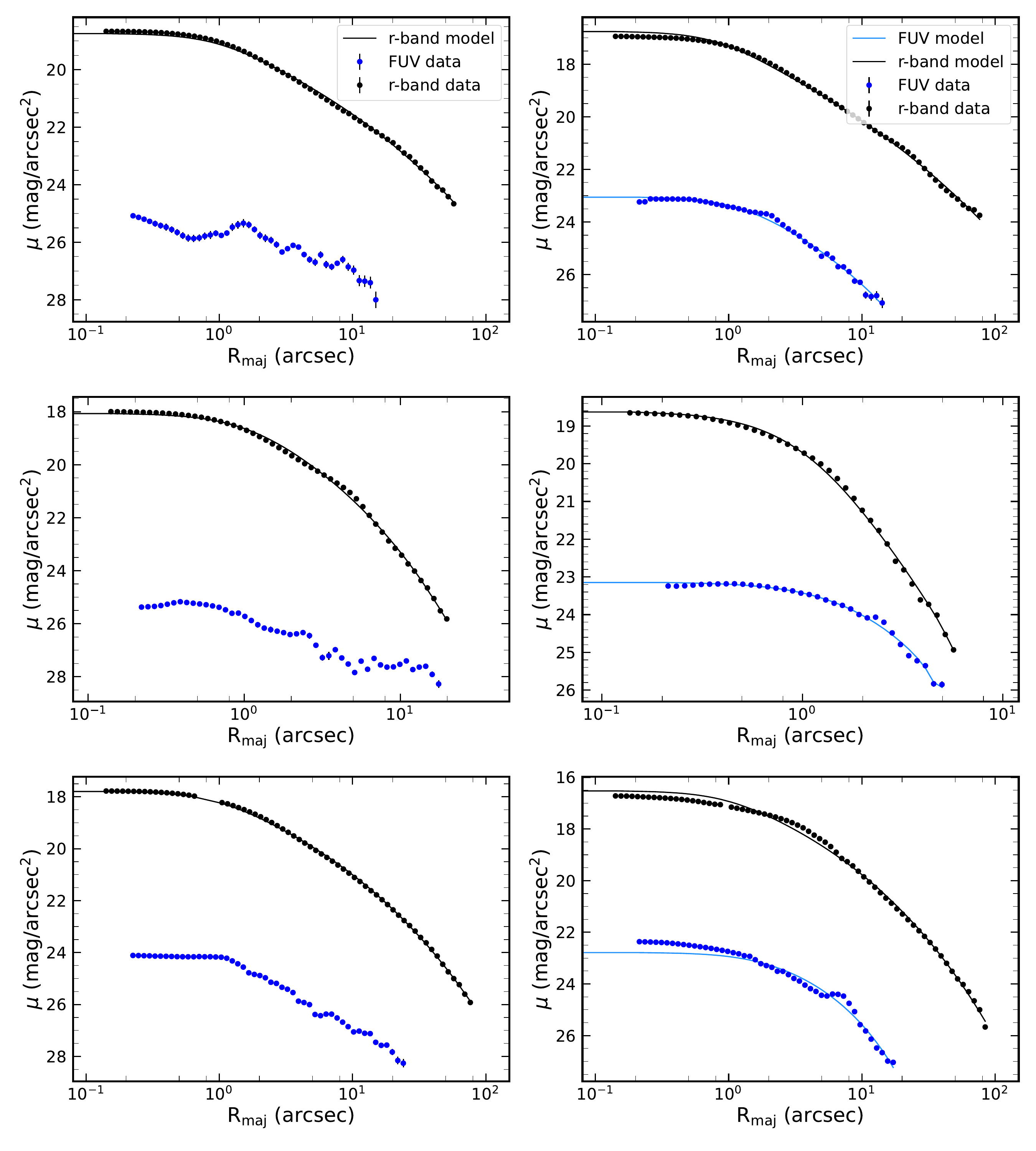}
    \caption{Surface brightness profiles for a subset of our ETGs, in the \decals $r$-band (\textcolor{black}{black} filled circles) and the UVIT FUV (blue filled circles). The 1-D S\'ersic fits to the $r$-band and FUV profiles are represented by the \textcolor{black}{blue} and black \textcolor{black}{curves,} respectively. The left-hand column shows examples of ETGs in which a S\'ersic function could not be fitted to the FUV surface brightness profiles because they are not smooth enough, while the right-hand column shows examples of galaxies where S\'ersic indices could be derived in both the FUV and $r$-band.}
    \label{fig:sersic_fitting}
\end{figure*}

\begin{comment}
\begin{figure*}
    \centering
    \includegraphics[width=0.99\linewidth]{sersic_fit_bad_jan.pdf}
    \vskip -15mm
    \includegraphics[width=0.99\linewidth]{sersic_fit_good_jan.pdf}
    \caption{Surface brightness profiles for a subset of our (six) ETGs, in the \decals $r$-band (red filled circles) and the FUV (blue filled circles). The 1-D S\'ersic fits to the $r$-band and FUV profiles are represented by black and blue solid lines respectively. The upper panel shows examples of ETGs in which a S\'ersic function could not be fitted to the FUV surface brightness profiles because they are not smooth enough, while the lower panel shows examples of galaxies where S\'ersic indices could be derived in both the FUV and $r$-band.}
    \label{fig:new_sersic}
\end{figure*}
\end{comment}

\begin{figure*}
    \centering
    \includegraphics[width=0.99\linewidth]{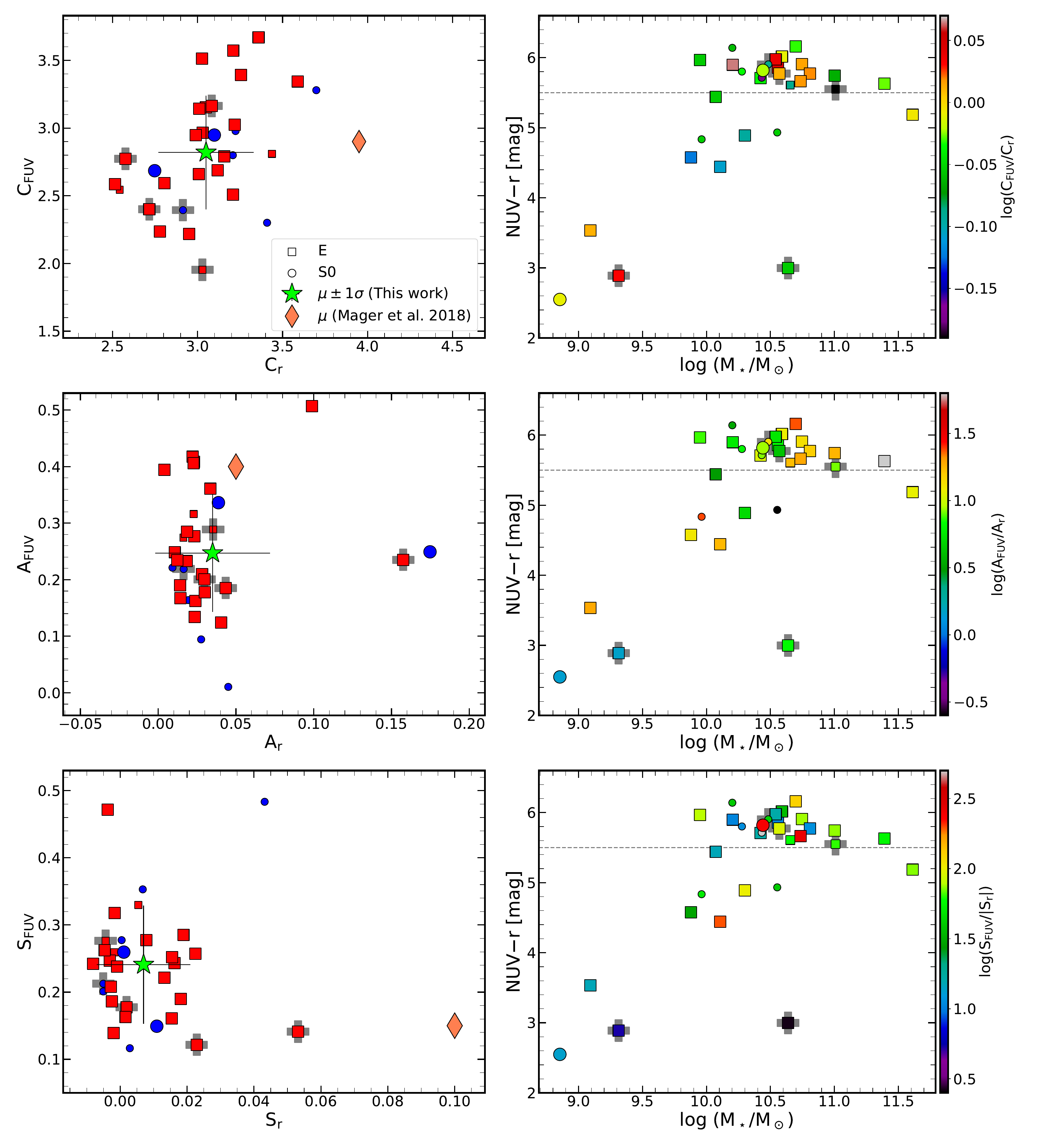}
    \caption{A comparison of the CAS parameters calculated from the FUV and $r$-band images of our ETGs. Elliptical (E) and lenticular (S0) galaxies are shown using different symbols. The parameters are measured within the same aperture in both bands (1.5 times the FUV petrosian radius). The larger markers indicate galaxies with SNR$_{\rm pixel}$ $\geq$ 2.5, while the smaller markers indicate galaxies with SNR$_{\rm pixel}$ between 2.0 and 2.5. The left-hand column plots the FUV vs $r$-band values of each parameter. \textcolor{black}{The mean value of each parameter from our work and \citet{Mager2018} is shown using the green star and orange diamond respectively. The error bar on the mean values from our study corresponds to 1$\sigma$ deviation in the distribution.} The right-hand column plots the GALEX-SDSS $(NUV-r)$ colour vs stellar mass, colour-coded by the log of the ratio between the FUV and $r$-band values of the parameter in question. Galaxies that are interacting are shown using grey crosses. \textcolor{black}{\textcolor{black}{Following \citet{Kaviraj2007}, a horizontal line separating star-forming and passive galaxies is indicated at $NUV-r$ = 5.5 mag.}}}
    \label{fig:CAS}
\end{figure*}

%%%%%%%%%%%%%%%%%%%%%%%%%%%%%%%%%%%%%%%%%%%%%%%%%%

\subsection{FUV and optical morphologies of ETGs are different - the UV emission in ETGs is driven by star formation}

\label{sec:starformation}

We proceed by comparing the quantitative morphologies of our ETGs in the optical and FUV using the CAS parameters calculated in the previous section. The left-hand column of Figure~\ref{fig:CAS} compares concentration (top row), asymmetry (middle row) and clumpiness (bottom row) in the FUV and optical wavelengths. Recall that, in each galaxy, the CAS parameters are measured in the same aperture (1.5 times the FUV petrosian radius) for both bands. The right-hand column plots the GALEX-SDSS $(NUV-r)$ colour vs stellar mass, with the galaxies colour-coded by the log of the ratio between the FUV and optical values of the parameter in question. Galaxies with FUV SNR$_{\rm pixel}$ $\geq$ 2.5 are shown using larger symbols, while those with FUV SNR$_{\rm pixel}$ between 2.0 and 2.5 are shown using smaller symbols. Interacting galaxies are indicated using grey crosses.  We note that the results do not change if the region within which these parameters are measured is not restricted to 1.5 times the FUV petrosian radius (see Figure~\ref{fig:cas_appendix} in the appendix). \textcolor{black}{Figure~\ref{fig:cas_appendix} is an alternate version of Figure~\ref{fig:CAS}, where the optical and FUV CAS parameters are calculated within 1.5 times of the respective petrosian radius of each band.} 

We use the $(NUV-r)$ colour in our analysis for consistency with the previous literature. Recall that $(NUV-r)$ serves as a useful tracer of the age of the dominant stellar population in galaxies \citep[e.g.][]{Yi2005,Kaviraj2007}. ETGs in which the presence of recent star formation is considered likely, based on integrated photometry, lie blueward of $(NUV-r) = 5.5$ \citep{Kaviraj2007}. ETGs with $(NUV-r)>5.5$, on the other hand, are relatively passive, and there is some ambiguity about whether the dominant UV sources in this galaxies are old or young. Evidently, our ETG sample spans the full spectrum of systems, from those that are clearly star-forming to those that are relatively passive. Indeed, at the higher end of our stellar mass range, the sample is dominated by galaxies redder than $(NUV-r)=5.5$. 

The mean values and standard deviations of $C_{\rm r}$, $A_{\rm r}$ and $S_{\rm r}$ are {\color{black} 3.05 $\pm$ 0.28, 0.035 $\pm$ 0.037 and 0.007 $\pm$ 0.014}, respectively, consistent with the optical CAS parameters of ETGs derived in the literature \citep[e.g.][]{Conselice2003, Lotz2004, 2008toledo,Hambleton2011, Mager2018}. In agreement with past work, our ETGs display negligible clumpiness and asymmetry in their optical images which are dominated by the old stellar populations that form the bulk of their stellar mass. In contrast, the mean values and standard deviations of $C_{\rm FUV}$, $A_{\rm FUV}$ and $S_{\rm FUV}$ are {\color{black} 2.82 $\pm$ 0.42, 0.247 $\pm$ 0.104 and 0.241 $\pm$ 0.085}, respectively. The values remain virtually unchanged if we restrict our ETGs to the subset with FUV SNR$_{\rm pixel}$ $\geq$ 2.5. While $C_{\rm r}$ is comparable or, in some cases, lower than $C_{\rm FUV}$, we find significant differences between the asymmetry and clumpiness parameters across the FUV and optical wavelengths. The values of $A_{\rm FUV}$ and $S_{\rm FUV}$ are always greater than their optical counterparts. Across our ETGs, the mean FUV-to-optical ratios of asymmetry and clumpiness are $\sim$13 and $\sim$82, respectively. %Across our ETGs, the mean FUV-to-optical ratios of asymmetry and clumpiness are \textbf{12.61} and \textbf{63.45}, respectively. 
\textcolor{black}{Note that a few of the bluest ETGs in our sample show relatively high values of $A_{\rm r}$ and $S_{\rm r}$, which results in lower UV-to-optical ratios (with the asymmetry ratio being less than 1 in two cases)}. Nevertheless, the asymmetry and clumpiness in the FUV are typically at least several factors (and in some cases several orders of magnitude) larger than their optical counterparts. It is worth noting that these trends are present irrespective of the stellar mass, $(NUV-r)$ colour, or the presence of an interaction in our ETGs. \textcolor{black}{In Figures \ref{fig:unsharp_decals} and \ref{fig:unsharp_fuv} we show unsharp-masked images in the \decals $r$-band and GALEX FUV for our ETGs. The number on each stamp corresponds to the ID of the galaxy in Table~\ref{tab:Tab1}. The unsharp-masking technique accentuates faint, low-contrast structures that may be embedded in a bright background. The figures qualitatively highlight the dissimilarity in clumpiness and asymmetry in the light distribution of ETGs in the two bands, mirroring the quantitative result derived above using the CAS parameters.}

Asymmetry and clumpiness (particularly the latter) trace structure in galaxy images. The significant differences seen in the CAS parameters in the UV and optical indicate that the light in these two wavebands do not originate from the same stellar populations in our ETGs. Since the optical light in ETGs is known to come from old stars, our results indicate that the UV is not driven by these old stellar populations. Furthermore, star formation is inherently patchy which causes galaxies with later type morphologies to have higher asymmetry and clumpiness in all wavelengths \citep[e.g.][]{Mager2018,Cheng2021}. The high asymmetry and clumpiness values in the UV demonstrate the presence of star formation in our ETGs. Note that most of the very massive (non-interacting) UV-red ETGs ($M_\star$ > 10$^{10.5}$ M$_{\odot}$) in our sample show $A_{\rm FUV}$/$A_{r}$ and $S_{\rm FUV}$/$S_{\rm r}$ values greater than 10, implying the presence of star-forming regions even in relatively passive ETGs. The fact that the structure is only seen in the UV and not in the optical confirms the suggestion in past work \citep[e.g.][]{Kaviraj2007} that the star formation in most ETGs is at a relatively low level, which affects the star formation sensitive UV image but not its optical counterpart. 

%These star-forming regions are preferentially spread around the centre of the massive galaxies in form of a faint UV halo or disk. \textbf{It's not clear how we can draw this specific conclusion about faint UV halos and a disk and where the star forming regions are in the galaxy. If we can't quantify this precisely best to remove this sentence.}

%{\color{teal} Finally, it is worth noting here that a fraction of galaxies sampled with UVIT lies in the vicinity of the Coma cluster \citep[Abell 1656, ][]{1989Abell}. The sample comprises close to 14 ETGs within a radius of 15 Mpc from the cluster center, including one of the brightest members of the cluster - NGC 4889. A previously done analysis by \citet{Yi_2011} investigated the possibility of sighting UV-upturn phenomena in luminous cluster galaxies. A majority of the galaxies present in the Coma cluster were found to be UV-weak (faint in UV) with no UV upturn galaxies. One of ETGs, IC 4045 is classfied as a UV upturn system \citep{2002Deharveng}}  
% \textbf{XXX} of our ETGs \textbf{(might be worth giving NGC numbers if they have them)} are classified as UV upturn systems \citep{Add the reference here}. 

Finally, it is worth noting that {\color{black} seven} ETGs in our sample (IC 3976, NGC 4875, NGC 4889, NGC 4883, NGC 4872, IC 4045 and LEDA 44656) are classified as `UV upturn' galaxies \citep{2002Deharveng, 2018Ali}, in which the UV flux is assumed to be driven exclusively by old stars. Such systems have been typically selected using combinations of UV, optical and mid-infrared colours \citep[e.g.][]{Yi_2011,2018Ali} which minimize the contribution of young stars to the UV. However, notwithstanding the assumption of old stars driving the UV flux in these system, the mean FUV-to-optical ratios of asymmetry ($\sim$9) and clumpiness ($\sim$93) are as high in UV upturn systems as it is in the rest of our ETG population. This suggests that the UV fluxes in these galaxies are also driven by (or at least significantly influenced by) recent star formation and not only by the old stellar population. Although the sample of such objects is relatively small, the large differences between the UV and optical morphologies in these UV upturn galaxies suggests that an assumption about old stars driving the UV flux in such systems cannot be made using integrated UV colours alone.

It is worth considering our results in the context of \citet{Mager2018}, who have performed the first systematic study of massive galaxies, of all morphological types, across the UV and optical wavelengths, using GALEX, HST and ground-based optical observations. Interestingly, the most substantial differences seen within a morphological class in Mager et al. are between the concentration and asymmetry of ETGs in the optical and UV wavelengths. In a similar vein to our results, they find that $C_{\rm FUV}$ and $A_{\rm FUV}$ in ETGs are higher than their optical counterparts and mirror that of spiral and irregular galaxies. While the authors do not arrive at a similar conclusion for the $S_{\rm FUV}$ parameter, this is likely due to the fact that the relatively low resolution of GALEX washes out small-scale structure in the UV images. In contrast, the much higher resolution of our UV images (which are well-matched to the resolution of the optical images) enables us to quantify the differences in both clumpiness and asymmetry in ETGs across the UV and optical wavelengths. \citet{Mager2018} conclude that the differences in the morphological parameters in ETGs between the UV and optical wavelengths can be explained by the presence of extended disks that contain recent star formation, consistent with the conclusions of our study. 

{\color{black} The mean values of the CAS parameters derived here are similar to those in the literature. For example, the mean values of $C_{\rm r}$, $A_{\rm r}$ and $S_{\rm r}$ in our study are around 3.05, 0.035, and 0.007, while they are 4.4, 0.02, and 0.00 and 3.95, 0.05, and 0.1 in \citet{Conselice_2003} and \citet{Mager2018} respectively. The values reported by \citet{Conselice2003} are slightly closer to our findings compared to \citet{Mager2018}. While our values of concentration and asymmetry in the optical are similar, \citet{Mager2018} find larger values of clumpiness. The two most likely causes of this difference are that a majority of the ETGs studied by \citet{Mager2018} are more nearby than those studied here and that the optical HST images used by \citet{Mager2018} have higher resolution than the \decals data used in this study. 

Similarly, the mean values of $C_{\rm FUV}$, $A_{\rm FUV}$ and $S_{\rm FUV}$ in \citet{Mager2018} are around 2.9, 0.4, 0.15 respectively, compared to 2.82, 0.25 and 0.24 in this study. Thus, while the values of concentration in the FUV are similar in both studies, the mean values of asymmetry and clumpiness are higher in our work compared to Mager et al. This is likely driven by the fact that our UVIT FUV images have significantly higher resolution, which enables them to identify more detailed structures in the UV than is possible using GALEX. We show the mean CAS parameters and standard deviations in the optical and FUV bands from our study and \citet{Mager2018} in the left-hand column of Figures~\ref{fig:CAS} and ~\ref{fig:cas_appendix}.}

\begin{figure}
    \centering
    \includegraphics[width=\columnwidth]{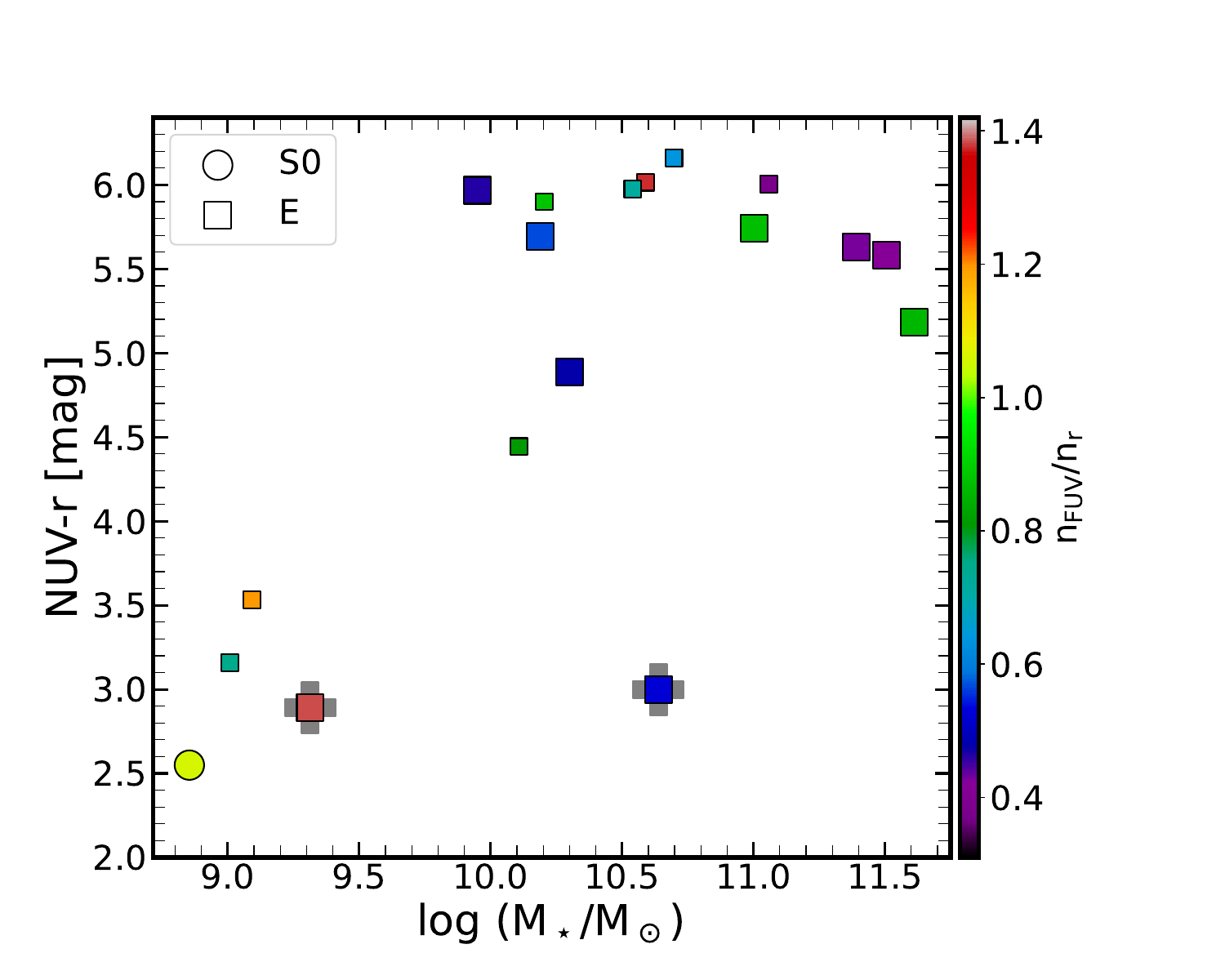}
    \caption{A comparison between the S\'ersic indices calculated from the FUV and $r$-band images of our ETGs. Galaxies that are ellipticals (E) and those that are lenticular (S0) are shown using different symbols. The larger symbols indicate galaxies where $\Delta n \leq 0.25$ (where $\Delta n$ is the error in $n$), while the smaller symbols indicate galaxies with 0.25 $< \Delta n \leq$ 1. The galaxies are colour-coded by the log of the ratio between the FUV and $r$-band S\'ersic indices. Galaxies that are interacting are shown using crosses.}
    \label{fig:sersic_cmp}
\end{figure}

We complete our study by comparing the S\'ersic indices of our ETGs in the UV and optical wavelengths. Figure \ref{fig:sersic_cmp} plots the $(NUV-r)$ colour vs stellar mass of our ETGs, colour-coded by the ratio of the S\'ersic indices in the UV and optical wavelengths. As noted in Section~\ref{sec:morphparams}, the lack of smoothness of the FUV profile, which is a direct result of the structure seen in the UV image, prevents us from fitting a S\'ersic function in the FUV to around 55 per cent of our ETGs. Therefore, the comparison can only be made for a subset of our galaxies (see numbers in Section~\ref{sec:morphparams}). The mean values of $n_{\rm FUV}$ and $n_{\rm r}$ for ETGs corresponding to $\Delta n$ $<$ 0.25 (1) are 2.06 (2.14) and 3.21 (3.15), respectively. In the vast majority of ETGs, the S\'ersic indices in the UV are lower than in the optical. The mean value and standard deviation of $n_{\rm{FUV}}$/$n_{\rm{r}}$ is 0.7 $\pm$ 0.3. In line with the conclusions from the morphological parameters above, this suggests that the UV light in our ETGs is more extended (or flatter) than in the optical, potentially signposting the presence of extended disks, as hypothesised by the study of \citet{Mager2018}. {\color{black}We note that a similar result has been reported in blue ETGs by \citet{Paspaliaris2023}.} 

% leads to the large error bars.   

%%%%%%%%%%%%%%%%%%%%%%%%%%%%%%%%%%%%%%%%%%%%%%%%%%

\section{Summary}
\label{sec:summary} 

The unexpected recent discovery of the ubiquitous presence of UV sources in nearby ETGs has challenged traditional models for their evolution, which postulated their formation in rapid starbursts at high redshift followed by passive ageing thereafter. The UV fluxes in many ETGs appear stronger than what can be produced via old stars alone, indicating the presence of recent star formation. The coincidence of blue UV-optical colours, gas and morphological disturbances out to at least intermediate redshift indicates the persistent presence of merger-driven star formation in ETGs at late epochs. Nevertheless, the possibility of the UV flux having contributions from both old and young stars, coupled with uncertainties in theoretical models, means that a definitive conclusion about the presence of star formation in all ETGs is difficult to achieve using integrated photometry alone. 

An unambiguous way of disentangling the source of the UV flux in ETGs is to look for the presence of structure in the UV images. Old stars dominate the optical images of ETGs, which are smooth, devoid of structure, exhibit low values of morphological parameters, such as asymmetry and clumpiness, and high values of the S\'ersic index. If the UV is also driven by old stars then the UV images will share the smoothness and lack of structure seen in the optical images. If, on the other hand, it is driven by young stars, then there should be significant structure in the UV images, which will manifest itself as higher values of asymmetry and clumpiness and lower values of the S\'ersic index than what is measured in the optical wavelengths. Crucially, this is true regardless of the strength of the UV flux, making this method a more effective discriminator between old and young-star driven UV flux than integrated photometry. 

Here, we have compared the UV and optical morphologies of a sample of 32 ETGs at {\color{black}$z<0.102$}, 93 per cent of which have redshifts less than 0.03. We have used deep UV and optical images from the DECam (via the \decals survey) and UVIT, which have similar resolution, which enables an `apples-to-apples' comparison. Furthermore, our ETGs lie in the part of the redshift vs stellar mass plane where galaxy populations are complete and unbiased, allowing us to probe the presence of star formation regardless of the strength of the UV flux in our ETGs. Our main results are as follows: 

\begin{itemize}

    \item Regardless of stellar mass, UV-optical colour or the presence of an interaction, the asymmetry and clumpiness of ETGs is typically significantly larger in the UV than in the optical. The mean FUV to optical ratios of asymmetry and clumpiness are $\sim$13 and $\sim$82 respectively. The asymmetry and clumpiness in the FUV are typically at least several factors (and in some cases several orders of magnitude) larger than in their optical counterparts. The significant amounts of UV structure indicates that the UV flux in \textit{all} ETGs in our sample is either dominated by, or has a significant contribution from, young stars. 

    \item  Interestingly, the trends above are also seen in seven ETGs which are classified as UV upturn systems in the literature i.e. galaxies in which the UV flux is assumed to be driven solely by old stars. This assumption appears incorrect, at least in these ETGs. It is plausible that the same may be true of other UV upturn systems, indicating that the assumption of old stars driving the UV flux in such systems cannot be made on the basis of integrated UV colours alone.  

    \item Mirroring the results from the morphological parameters, the S\'ersic indices of ETGs are lower in the UV than in the optical, suggesting the presence of disky structures. Furthermore, unlike in the optical, the significant UV structure produces FUV radial profiles which are not smooth, making it difficult to even fit S\'ersic parameters to around 55 per cent of our ETGs. 

\end{itemize}
Our results provide independent evidence and important corroboration for recent work that suggests the presence of star formation in ETGs. This has been based both on the widespread presence of UV sources in local ETGs and the fact that the distribution of their UV-optical colours appears unchanged out to intermediate redshift (where the Universe is too young for old stars to be in place). The ubiquitous presence of structure in the FUV images of our ETGs, regardless of stellar mass, UV-optical colour and the presence of interactions, indicates that \textit{all} ETGs, including those on the UV red sequence, host star formation.

%%%%%%%%%%%%%%%%%%%%%%%%%%%%%%%%%%%%%%%%%%%%%%%%%%

\section*{Acknowledgements}
\textcolor{black}{We thank the referee for many constructive comments which improved the quality of the original manuscript.} DP thanks Sonika Piridi for assistance in UVIT data reduction. DP also acknowledges the Inter University Centre for Astronomy and Astrophysics (IUCAA), Pune, India for providing hospitality and travel grants. SK acknowledges support from the STFC [grant number ST/X001318/1] and a Senior Research Fellowship from Worcester College Oxford. KS acknowledges support from the Indian Space Research Organisation (ISRO) funding under project PAO/REF/CP167. For the purpose of open access, the authors have applied a Creative Commons Attribution (CC BY) licence to any Author Accepted Manuscript version arising from this submission. 

%%%%%%%%%%%%%%%%%%%%%%%%%%%%%%%%%%%%%%%%%%%%%%%%%%

\section*{Data Availability}
The GALEX-SDSS-WISE Legacy Catalog can be found at: \url{https://salims.pages.iu.edu/gswlc/}. Data from the Galaxy Zoo citizen-science project is located at: \url{https://data.galaxyzoo.org/}. This publication uses archival data from the \textit{AstroSat} mission of the Indian Space Research Organisation (ISRO), which is archived at the Indian Space Science Data Centre (ISSDC) and can be found here: \url{https://astrobrowse.issdc.gov.in/astro\_archive/archive/Home.jsp}. Data from the \decals survey is located at: \url{https://www.legacysurvey.org/decamls/}.



\bibliographystyle{mnras}
\bibliography{references}

%\begin{thebibliography}{99}
%\end{thebibliography}

%%%%%%%%%%%%%%%%%%%%%%%%%%%%%%%%%%%%%%%%%%%%%%%%%%

%%%%%%%%%%%%%%%%% APPENDICES %%%%%%%%%%%%%%%%%%%%%

\appendix

\section{UV and optical morphological parameters not restricted to the same region of the galaxy}

The analysis presented in the sections above employs morphological parameters in the UV and optical that are derived within 1.5 times the petrosian radius in the FUV image. Restricting the calculation to the same region of the galaxy enables us to perform a like-with-like comparison of UV and optical morphology. In this section we present, in Figure \ref{fig:cas_appendix}, a comparison of UV and optical morphology without this restriction. Here, the UV and optical parameters are derived within 1.5 times the petrosian radius of the image in question. We find that the general trends in the UV and optical CAS parameters remain unchanged. Our results are therefore not sensitive to the region of the galaxy within which the CAS parameters are calculated in the UV and optical wavelengths. 

\begin{figure*}
    \centering
    \includegraphics[width=\linewidth]{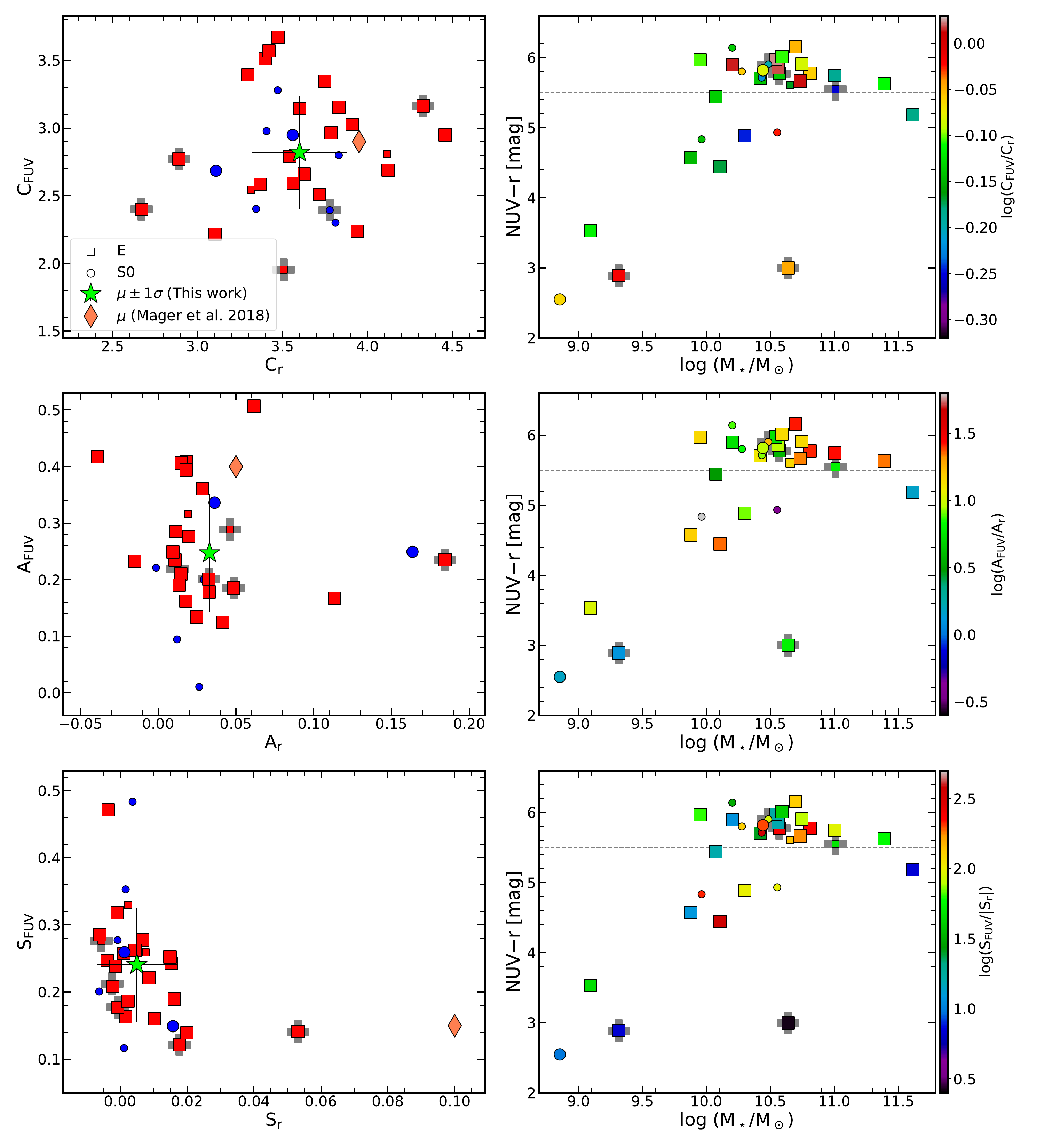}
    \caption{A comparison between the CAS parameters calculated from FUV and $r$-band images of our ETGs. Galaxies that are ellipticals (E) and those that are lenticular (S0) are shown using different symbols. Unlike Figure \ref{fig:CAS}, the parameter measurements in the UV and optical are not restricted to the same apertures in both bands. The larger symbols indicate galaxies with SNR$_{\rm pixel}$ $>$ 2.5, while the smaller symbols indicate galaxies with SNR$_{\rm pixel}$ between 2.0 and 2.5. The left-hand column plots the FUV vs $r$-band values of each parameter. \textcolor{black}{The mean value of each parameter from our work and \citet{Mager2018} is shown with green star and orange diamond markers respectively. The error bar on the mean values from our study corresponds to 1$\sigma$ deviation in the distribution.} The right-hand column plots the GALEX-SDSS $(NUV-r)$ colour vs stellar mass colour coded by the log of the ratio between the FUV and $r$-band values of the parameter in question. Galaxies that are interacting are shown using grey crosses. \textcolor{black}{Following \citet{Kaviraj2007}, a horizontal line separating star-forming and passive galaxies is indicated at $NUV-r$ = 5.5 mag.}}
    \label{fig:cas_appendix}
\end{figure*}

\begin{figure*}
    \centering
    \includegraphics[width=\linewidth]{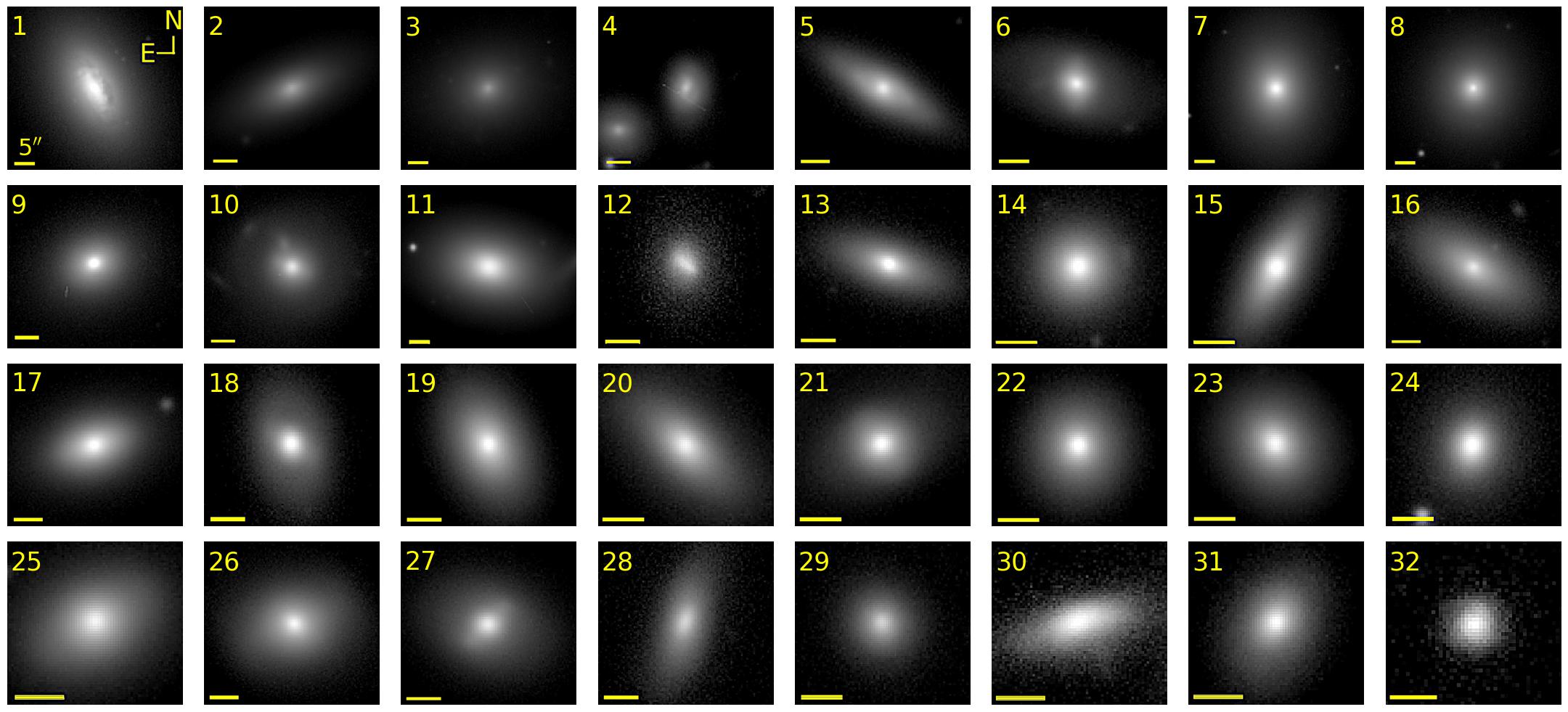}
    \caption{\textcolor{black}{Unsharp-masked \decals $r$-band images for all 32 ETGs in our final sample. The galaxy IDs from Table~\ref{tab:Tab1} are indicated in the upper left-hand side of each stamp.}}
    \label{fig:unsharp_decals}
\end{figure*}
\begin{figure*}
    \centering
    \includegraphics[width=\linewidth]{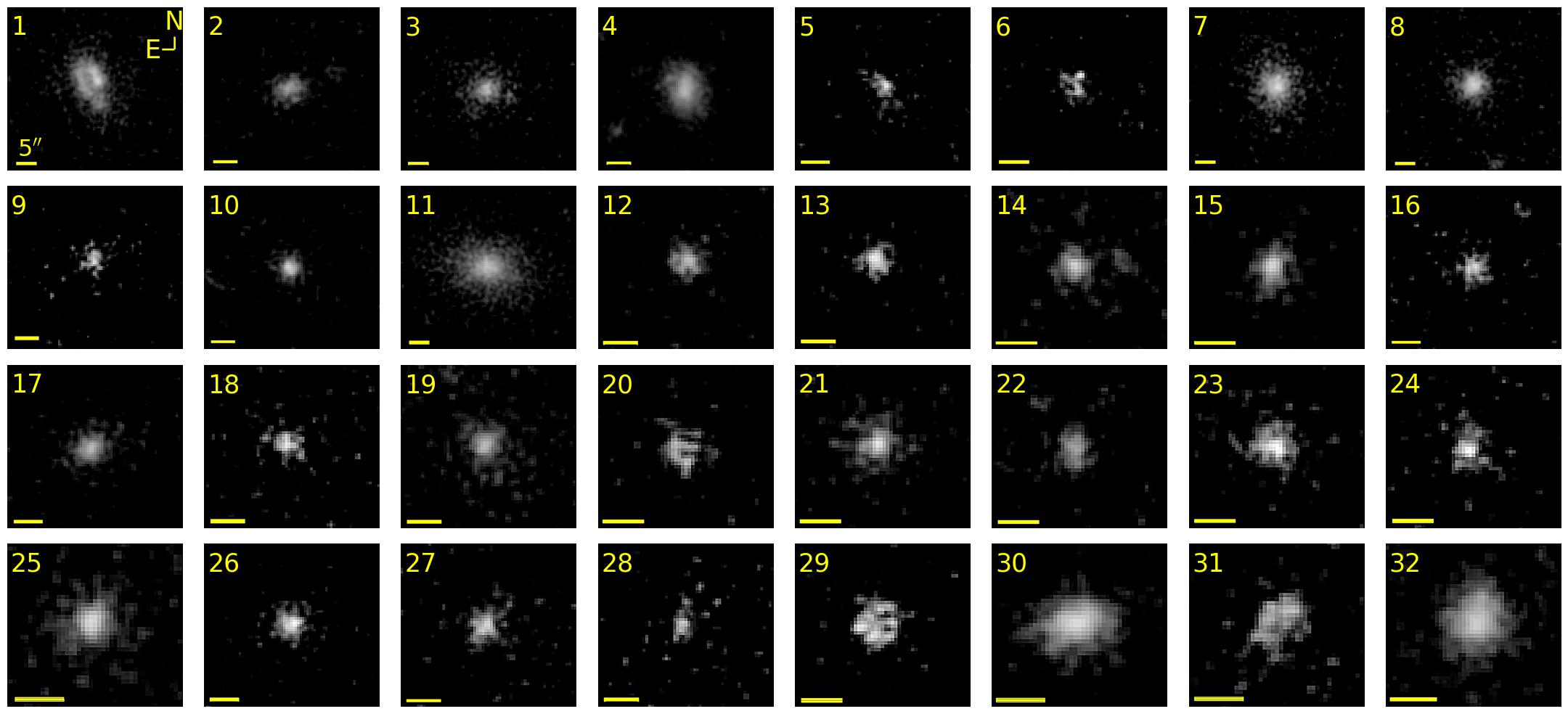}
    \caption{\textcolor{black}{Unsharp-masked FUV images fo all 32 ETGs in our sample. The galaxy IDs from Table~\ref{tab:Tab1} are indicated in the upper left-hand side of each stamp. The region shown around each galaxy matches that of their optical counterpart image in Figure~\ref{fig:unsharp_decals}. }}
    \label{fig:unsharp_fuv}
\end{figure*}

%%%%%%%%%%%%%%%%%%%%%%%%%%%%%%%%%%%%%%%%%%%%%%%%%%

% Don't change these lines
\bsp	% typesetting comment
\label{lastpage}
\end{document}